\RequirePackage{fix-cm}
\documentclass[envcountsect,numbook]{svjour3_arxiv4}   

\renewcommand{\PACSname}{\textbf{JEL classification}\enspace}

\usepackage{mathtools,mathptmx}
\mathtoolsset{showonlyrefs}
\usepackage{amssymb,amsmath,amsfonts,latexsym} 
\usepackage{booktabs,tabularx,ltablex}
\usepackage{hyperref}
\usepackage{url}
\usepackage[utf8]{inputenc}
\usepackage{bbm,dsfont,textcomp}
\usepackage[normalem]{ulem}

\usepackage{graphicx,rotating,pdflscape,placeins,afterpage}
\usepackage{caption,float,subcaption}
\usepackage{float}
\graphicspath{{./}}

\usepackage[12pt]{extsizes}
\usepackage{geometry}
\usepackage{ifthen}
\usepackage{calrsfs}
\DeclareMathAlphabet{\pazocal}{OMS}{zplm}{m}{n}
\let\mathcal\pazocal
\allowdisplaybreaks  

\newtheorem{prop}{Proposition}[section]
\newtheorem{lem}{Lemma}[section]
\newtheorem{ass}{Assumption}[section]
\newtheorem{bem}{Remark}[section]
\smartqed

\numberwithin{equation}{section}
\numberwithin{table}{section}
\numberwithin{figure}{section}

\usepackage{dcolumn}
\newcolumntype{d}{D{.}{.}{4.1}}
\newcolumntype{e}{D{.}{.}{2.2}}
\newcommand\mc[1]{\multicolumn{1}{r}{#1}}
\newcommand\mcl[1]{\multicolumn{1}{r|}{#1}}

\makeatletter
\def\DC@endright{$\hfil\egroup\@dcolcolor\box\z@\box\tw@\dcolreset}
\def\dcolcolor#1{\gdef\@dcolcolor{\color{#1}}}
\def\dcolreset{\dcolcolor{black}}
\dcolcolor{black}
\makeatother

\usepackage{natbib} 
\usepackage{enumitem,xcolor}
\definecolor{myblue}{rgb}{0.0,0.0,1.0}
\definecolor{myred}{rgb}{0.8,0,0} 
\definecolor{mygreen}{rgb}{0,0.7,0}
\definecolor{mygray}{rgb}{0.5,0.5,0.5}
\newcommand{\neu}{\color{black}}

\usepackage[]{todonotes} 
\definecolor{markierung}{rgb}{0.7,1,0.7} 
\usepackage{soul}
\sethlcolor{markierung}
\makeatletter
\if@todonotes@disabled

\else

\fi
\makeatother

\def \R{\mathbb{R}}               
\def \N{\mathbb{N}}               
  
\def \P{\mathbb{P}}  
\def \E{\mathbb{E}}  
\def \stratc{\Psi}
\def \stratd{\Phi}
\def \scalefac{\gamma}
\def \Vini{v}
\def \rebalcost{\mathcal{D}}
\def \rebalcosts{\rebalcost_T}
\def \transactcost{\mathcal{L}}
\def \transactcosts{\transactcost_T}
\def \hsalop{\mathcal{H}}
\def \hshir{\mathcal{G}}
\def \Hminindex{\mathcal{I}_{\min}}
\def \arbbound{\underline{v}} 
\newcommand{\Ito}{{Itô}}
\def \one{\mathbbm{1}}
\def \gauss{Z}
\newcommand{\Varianz}{\operatorname{Var}}
\newcommand{\Var}[1]{\Varianz(#1) }

\newcommand{\Cov}{\operatorname{Cov}}

\begin{document}
	\title{Discretization of continuous-time arbitrage strategies in financial markets with fractional Brownian motion}
	
	\titlerunning{Arbitrage strategies in markets with fractional Brownian motion}        
	
	\author{ Kerstin Lamert \and Benjamin R. Auer  \and Ralf Wunderlich}
	\authorrunning{K.~Lamert, B.R.~Auer  and R.~Wunderlich}
	
	\institute{
		Kerstin Lamert \at 
		Brandenburg University of Technology Cottbus-Senftenberg, Institute of Mathematics, Platz der Deutschen Einheit 1, 03046 Cottbus, Germany;  
		\email{\texttt{kerstin.lamert@b-tu.de}} 
		\and
		Benjamin R. Auer \at
		Friedrich Schiller University Jena, Chair of Finance, Carl-Zeiss-Str. 3, 07743 Jena, Germany;    
		\email{\texttt{benjamin.auer@uni-jena.de}}           
		\and
		Ralf Wunderlich \at
		Brandenburg University of Technology Cottbus-Senftenberg, Institute of Mathematics, Platz der Deutschen Einheit 1, 03046 Cottbus, Germany;  
		\email{\texttt{ralf.wunderlich@b-tu.de}} 
	}
	
	\date{Version of  \today}
	
	\maketitle	
	
	\begin{abstract}
		This study evaluates the practical usefulness of continuous-time arbitrage strategies designed to exploit serial correlation in fractional financial markets. Specifically, we revisit the strategies of \cite{Shiryaev1998} and \cite{Salopek1998} and transfer them to a real-world setting by distretizing their dynamics and introducing transaction costs. {\neu In Monte Carlo simulations with various market and trading parameter settings as well as a formal analysis of discretization error, we show that both are promising with respect to terminal portfolio values and loss probabilities. These features and complementary sparsity make them worth serious consideration in the toolkit of quantitative investors.}
		
		\renewcommand{\PACSname}{\textbf{JEL classification}\enspace}
	\end{abstract}
	
	\renewcommand{\thefootnote}{\arabic{footnote}}
	
	\section{Introduction}
	\label{sec:intr}
	Motivated by the challenge they pose to the traditional notion of efficient capital markets, financial research has intensively studied investment strategies which solely rely on past asset price information \cite[see][]{Goyal2018}. Among the most prominent studies, \cite{Jegadeesh1993,Jegadeesh2001} have shown that cross-sectional momentum, i.e., buying past winners and selling past losers, is highly beneficial.\footnote{For the identification of winners and losers, relative past performance can be quantified via cumulative returns or established reward-to-risk performance measures \cite[see][]{Rachev2007}.} In addition, \cite{Moskowitz2012} identify a time-series momentum effect according to which single assets exhibit exploitable trending behavior.\footnote{\cite{Marshall2017} establish a connection between time-series momentum and the popular moving average trading rules of \cite{Brock1992}. {\neu \cite{Zakamulin2024} uncover the price processes under which more general moving average rules perform best.}} What these strategies have in common is that their profitability is linked to a positive serial correlation in asset price movements \cite[see][]{Pan2004,Hong2015}.
	
	Even though momentum investing has become a standard in the mutual fund industry \cite[see][]{Barroso2015}, financial research and practice has paid surprisingly little attention to a very interesting strand of mathematical literature developing arbitrage strategies for assets with serially correlated returns. It is well known that \textit{pure arbitrage}, i.e., the realization of risk-less profits from zero initial investment, is impossible in a traditional Black-Scholes market with standard Brownian motion (sBm). In contrast, arbitrage opportunities can arise in markets where asset prices are driven by a fractional Brownian motion (fBm) which dates back to \cite{Mandelbrot1968} and superimposes memory features on asset returns. In a continuous-time setup with slowly decaying positive serial correlation, i.e., the fractional Black-Scholes model of \cite{Cutland1995}, the theoretical studies of \cite{Shiryaev1998} and \cite{Salopek1998} show that risk-less profits can be earned by buying high-priced and short-selling low-priced assets in adequate numbers. \cite{Bayraktar2005} extend the work of \cite{Shiryaev1998} by incorporating stochastic volatility. \cite{Rogers1997} and \cite{Cheridito2003} develop additional but more complex strategies.
	
	While the simplicity of the Shiryaev and Salopek arbitrage strategies and empirical evidence on memory in equity, futures and fund returns \cite[see][]{Wang2012,DiCecare2015,Coakley2016} make them appealing for investment practice, they are built on the premise of continuous-time trading with no frictions. \cite{Cheridito2003} and \cite{Guasoni2006} highlight that, in a fractional Black-Scholes world, arbitrage opportunities vanish with the introduction of a minimal waiting time between subsequent transactions, i.e.,~discrete-time trading, and proportional transaction costs of any positive size, respectively.\footnote{{\neu \cite{Peyre2017} shows that certain simple strategies (i.e., linear combinations of buy-and-hold investments with regular rebalancing) do not generate any arbitrage profits even if there is no minimal waiting time between trades.} For research on the role of other frictions, see \cite{Guasoni2019,Guasoni2021}.} However, this does not necessarily mean that the above strategies should be discarded. When suitably discretized and parameterized, they may not be entirely self-financing and risk-free, but still provide positive expected payoffs at a low risk of loss. In other words, they could share some valuable properties with \textit{statistical arbitrage} strategies \cite[see][]{Bondarenko2003,Luetkebohmert2019}. {\neu Just recently, two interesting strategies of this kind have been proposed to exploit fBm market behavior. \cite{Garcin2022} enters (long or short) trading positions by explicitly forecasting future realizations of the fBm, whereas \cite{Xiang2024} construct a (long-only) buy-and-hold strategy with growing stop-profit boundary. However, both strategies involve tuning a variety of para\-meters. In addition, the former requires more than just current prices to derive an investment decision and the latter only benefits from rising asset prices. As we will see, these are limitations compared to the simpler yet more flexible strategies we focus on.}
	
	The goal of our study is straightforward. After exploring the properties and the economic intuition of the Shiryaev and Salopek strategies, the core objective of our study is to investigate their investment performance in a real-world setting. This means that, in a first step, we discretize the strategies and install different forms of transaction costs. This is not trivial because discretization alone makes the strategies lose their self-financing property and requires suitable countermeasures to maintain tradeability. In a second step, we perform an extensive Monte Carlo study for the discretized versions of the strategies. Here, we are particularly interested in whether they deliver positive terminal portfolio values on average and display acceptably small loss probabilities. We focus on these two quantities because they are central to established arbitrage definitions and allow a modern downside-oriented investment evaluation \cite[see][]{Eling2007,Cumova2014}. {\neu To answer our research question, we use the circulant embedding method of \cite{Wood1994} for fBm simulation which, in contrast to alternatives, preserves the basic features of fBms and shines by remarkable computation speed \cite[see][]{Coeurjolly2000,Kijiama13}.} We analyze the strategies with asset and trading parameters tailored to the current market environment exhibiting, for example, significantly falling transaction costs \cite[see][]{Chordia2014}. Furthermore, we conduct a variety of sensitivity checks to identify the situations in which they perform best and worst. This and {\neu a supplementary formal inspection of discretization error} result in an intuitive guide on how to chose, for example, the ideal candidate assets, parameters and trading frequencies of the strategies.
	
	The remainder of our study is organized as follows. Section \ref{sec:method} introduces the fractional Black-Scholes model, discusses the corresponding Shiryaev and Salopek arbitrage strategies and translates them to a discrete-time setting with transaction costs. Section \ref{sec:Analysis} presents our Monte Carlo study examining the impact of discretization, transaction costs, model parameters as well as trading horizon and frequency on the strategies. {\neu Section \ref{sec:emp} illustrates empirical implementation issues by applying them to an exemplary data set of emerging market stocks.} Section \ref{sec:concl} concludes and outlines directions for future research.
	
	\section{Theoretical framework} \label{sec:method}
	\subsection{Continuous-time market setup}\label{subsec:ContinuousModel}
	We start our analysis by specifying the asset price behavior in a fractional Black-Scholes model and explain how self-financing portfolios are formed in such an environment.
	
	\paragraph{Asset prices}
	For a fixed date or investment horizon $T>0$, we consider a filtered probability space $(\Omega,\mathcal{F},\mathbb{F},\P)$ with
	standard filtration $\mathbb{F}=(\mathcal {F}_t)_{t \in [0,T]}$ and assume that all processes are $\mathbb{F}$-adapted. In this context, the \textit{fractional Black-Scholes model} suggests that we have one risk-free asset with constant price $S^0_t=1$
	and $d$ risky assets with a price process $S=(S^1,\ldots,S^d)$ defined on $[0,T]$ by the stochastic differential equations (SDEs)
	\begin{align}\label{eq:differentialequ}
		dS_t^i=\mu^iS_t^idt+\sigma^iS_t^idB_t^{H^i}, \quad S_0^i=s_0^i, \quad i=1,\ldots,d.
	\end{align}
	Here, the drifts or expected returns $\mu^i\in\mathbb{R}$, volatilities $\sigma^i>0$ and initial prices $s_0^i>0$ are given constants. In contrast to the standard Black-Scholes model, the SDEs are not driven by sBms (or Wiener processes) but by fBms $B_t^{H^i}$ with Hurst parameters $H^i\in(0.5,1)$. The fBms are assumed to be independent, which is reasonable when the risky assets are, for example, certain types of industry portfolios, investment funds or commodity futures baskets \cite[see][]{Erb2006,Badrinath2011}.\footnote{A formal discussion of correlated fBms can be found in \cite{Amblard2013}.}
	
	The one-dimensional fBm $(B_t^H)_{t\in[0,T]}$ is a centered Gaussian process with covariance 
	\begin{align}
		\label{fBm_covariance}
		\textrm{Cov}(B_t^H,B_s^H)=\frac{1}{2}\left(|t|^{2H}+|s|^{2H}-|t-s|^{2H}\right)\quad  \text{for} \quad H\in(0,1).
	\end{align}
	It is a self-similar process with stationary increments. The special case $H=0.5$ reflects the sBm which has independent increments and enjoys the martingale property. For $H\neq 0.5$, the increments are correlated and the process is not a martingale. This implies memory effects for the asset returns $dS^i_t/S^i_t$. For the range $H\in(0.5,1)$ considered in the fractional Black-Scholes world, asset returns are positively correlated, i.e., persistent. While, for $H=0.5$, we have to rely on \Ito's calculus, $H>0.5$ allows us to define stochastic integrals w.r.t.~$B_t^{H}$ path-wise in the Riemann-Stieltjes sense \cite[see][]{Sottinen2003}. Thus, the solutions of the SDEs in \eqref{eq:differentialequ} are
	\begin{align}
		S_t^i=s_0^i\, \exp\big\{\mu^i t+\sigma^i B_t^{H^i}\big\}, \quad  t \in \left[0,T\right]. \label{eq:BlackScholes}
	\end{align}
	The processes defined in \eqref{eq:BlackScholes} are called geometric fBms.\footnote{For the sBm with $H^i=0.5$, we have $S_t^i=s_0^i\, \exp\{(\mu^i-(\sigma^i)^2/2) t+\sigma^i B_t^{H^i}\}$. Properties of fBms and the associated integration theory are outlined in \cite{Biagini2008}, \cite{Mishura2008} and \cite{Bender2011}.} 
	
	\paragraph{Portfolios}
	Asset transactions in a fractional Black-Scholes market are modeled based on the usual assumptions of permanent trading, unlimited borrowing and short-selling, real asset quantities and contemporary agreement of buying and selling prices. {\neu Furthermore, we assume that trades do not affect market prices and, at this point, there are no transaction costs, fees or taxes.} We can describe the trading activity of an investor by an initial amount of capital $\Vini\in\R$ and a $\mathbb{F}$-adapted process 
	\begin{align}
		\label{eq:strat_cont}
		\stratc=(\stratc_t)_{t\in\left[0,T\right]}=(\Psi_t^0,\Psi_t^1,\ldots,\Psi_t^d)_{t\in\left[0,T\right]}.
	\end{align}
	Here, $\Psi_t^0\in\mathbb{R}$ and $\Psi_t^1,\ldots, \Psi_t^d\in\mathbb{R}$ denote the numbers of risk-free and risky assets held by the investor at time $t$, respectively. $\Psi$ is called \textit{portfolio} or \textit{trading strategy}. The investor has a long (short) position if $\Psi_{t}^i, i=0,\ldots,d,$ is positive (negative). At time $t$, the value $V_t^{\stratc}$ of the portfolio $\stratc$ is given by
	\begin{align}
		V_t^{\stratc} = \sum_{i=0}^{d} \Psi_t^iS_t^i.\label{eq:continuousValue}
	\end{align}  
	$V^\Psi=(V_t^{\stratc})_{t\in[0,T]}$ is the \textit{value process} of $\Psi$. In an arbitrage context, $\Psi$ is assumed to be \textit{self-financing}. This means that there is no exogenous infusion or withdrawal of capital after the purchase of the portfolio. Rebalancing the portfolio must be financed solely by trading the  $d+1$ available assets. Mathematically, self-financing in a continuous-time market is defined by the property that the value process for all $t\in[0,T]$ can be expressed as 
	\begin{align}
		\label{eq:SelfFinancing}
		V_t^{\stratc} = 	\Vini + G_t^{\stratc}, \quad\text{where }~~G_t^{\stratc}=\sum_{i=0}^{d} \int_{0}^{t}\Psi_s^idS^i_s.
	\end{align}
	$G^\Psi=(G_t^{\stratc})_{t\in[0,T]}$ is the \textit{gain process} of $\Psi$, where the gain $G_t^{\stratc}$ at time $t$ is given by a sum of stochastic integrals w.r.t. geometric fBms \cite[see][]{Salopek1998}.\footnote{{\neu Here, it is crucial to apply forward or Riemann-Stieltjes integrals because alternative stochastic integral definitions contradict economic intuition \cite[see][]{Hu2003,Rostek2013}.}} The self-financing condition \eqref{eq:SelfFinancing} can also be stated in differential form, i.e., $dV_t^{\stratc}=\sum_{i=0}^{d} \Psi_s^idS^i_s$. It shows that, for a self-financing strategy, the changes in portfolio value are not due to rebalancing but rather to changes in asset prices.
	
	\subsection{Continuous-time arbitrage}
	\label{sec:ArbitrageStrategies}
	In a standard Black-Scholes world, arbitrage is impossible. This is a consequence of the fundamental theorem of asset pricing \cite[see][]{Delbaen1994,Bjoerk2003} and has its roots in the sBm martingale property. In contrast, a fBm with $H\neq 0.5$ behaves predictably such that our fractional Black-Scholes world offers arbitrage opportunities which can be exploited by constructing suitable arbitrage portfolios. A self-financing portfolio $\Psi$ is called an \textit{arbitrage portfolio} if its value process satisfies
	\begin{equation}\label{eq:arbitrage}
		\begin{aligned}
			&\text{(i)}& & V_0^{\stratc}=0, \\
			&\text{(ii)}& & \P(V_t^{\stratc} \ge {\neu -\arbbound} \text{ for all } t\in(0,T])=1, ~~{\neu \text{for some constant } \arbbound\ge 0,} 	\\  
			&\text{(iii)}& & {\neu \P(V_T^{\stratc}\ge0)=1}~\text{ and }~ \P(V_T^{\stratc}>0)>0.
		\end{aligned}
	\end{equation}
	This implies that arbitrage is essentially the possibility to generate a positive amount of money without having to invest any initial capital and without any risk of loss. 
	{\neu The value of the arbitrage portfolio is bounded from below at all times and non-negative at the end of the investment horizon.}\footnote{\neu A negative lower bound rules out so-called doubling-type arbitrage strategies. Portfolios satisfying (i) to (iii) are also referred to as nds-admissible (no doubling strategies) arbitrages \cite[see][]{Bender2011}.}
	In the following, we present two simple arbitrage strategies satisfying all three properties in \eqref{eq:arbitrage} {\neu with $\arbbound=0$. That is, their portfolio values never fall below zero.}
	
	\begin{bem}
		\label{rem_scaling}	
		For every arbitrage strategy $\stratc$, the scaled strategy $\widetilde \stratc=\scalefac \stratc$ with $\scalefac>0$ is also an arbitrage strategy. This is because \eqref{eq:continuousValue} and \eqref{eq:SelfFinancing} imply $V^{\widetilde \stratc} = \scalefac V^{ \stratc}$ and $G^{\widetilde \stratc} = \scalefac G^{\stratc}$, respectively. Since the initial value $V^{ \stratc}_0$ is zero, we also have $V^{\widetilde \stratc}_0=0$. Hence, $\widetilde \stratc$ satisfies  the self-financing condition $V_t^{\widetilde \stratc} = v + G_t^{\widetilde \stratc}$ in \eqref{eq:SelfFinancing} with $v=0$. Finally, $V^{\widetilde \stratc} = \scalefac V^{ \stratc}$ implies that  $\widetilde \stratc$ fulfills  the arbitrage conditions in \eqref{eq:arbitrage}.
	\end{bem}
	
	\paragraph{Shiryaev strategy}
	\cite{Shiryaev1998} proposes a strategy that generates an arbitrage portfolio consisting of the risk-free asset and one risky asset.\footnote{{\neu In an independent study, the same strategy has been derived by \cite{Dasgupta2000}.}} Thus, we have $d=1$. For simplicity, we denote the drift, volatility and Hurst coefficient of the risky asset by $\mu$, $\sigma$ and $H$, respectively. For the two assets, the strategy suggests entering the time $t\in[0,T]$ positions 
	\begin{align}
		\label{eq:Shiryaev_stratc}
		\begin{array}{rcccl}
			\Psi_{t}^0&=&\frac{1}{s_0^1} \big({(s_{0}^1})^2-(S_t^1)^2\big) &=&s_0^1\big(1-\exp\big\{{2\mu t+2\sigma B_{t}^H}\big\}\big),\\ [1ex]
			\Psi_{t}^1&=& \frac{2}{s_0^1} ({S_{t}^1}-{s_0^1})&=&2\big(\exp\big\{\mu {t}+ \sigma B_{t}^H\}-1\big).
		\end{array}
	\end{align}
	At every $t$, it compares the value $S_t^1$ of a pure risky investment with an alternative investment of the initial risky asset price $s_0^1$ in the risk-free asset. Because, in our market model, we have $S_t^0=1$, this alternative investment has a constant value of $s_0^1$.\footnote{The strategy can be easily generalized to markets with risk-free asset prices $S_t^0=e^{rt}$, where $r\ge 0$ denotes the risk-free rate \cite[see][]{Shiryaev1998}.} If the value $S_t^1$ of the pure risky investment exceeds (falls below) the value $s_0^1$ of the alternative investment, the investor holds a long (short) position in the risky asset and a short (long) position in the risk-free asset. In the case of equality, he is not invested and the portfolio value is zero. The number of risky asset shares $\Psi_{t}^1$ in \eqref{eq:Shiryaev_stratc} does not depend on the initial risky asset price $s_0^1$ but on the parameters $\mu$, $\sigma$ and $H$. 
	
	\cite{Shiryaev1998} shows that the strategy $\stratc$ in \eqref{eq:Shiryaev_stratc} is self-financing. Furthermore, at time $t=0$, we have $\Psi_0^0=\Psi_0^1=0$ and hence $V_0^\stratc=0$, i.e., no initial investment is required. Substituting \eqref{eq:Shiryaev_stratc} into \eqref{eq:continuousValue} shows that the portfolio value at any time $t\in[0,T]$ is  
	\begin{align}
		\label{eq:TermValueShiraev}
		V^\Psi_t=(S_t^1-s_0^1)^2/s_0^1 \ge 0.
	\end{align}
	{\neu Thus, condition (ii) in \eqref{eq:arbitrage}  holds with the lower bound $\arbbound=0$. We also obtain $\P(V_T^{\stratc}\ge0)=1$ and
		$\P(V_T^{\stratc}>0)>0$.} This shows that $\stratc$ is indeed an arbitrage strategy.
	
	{\neu 
		Relation \eqref{eq:TermValueShiraev} and the fact that the asset price $S_t^1$ is lognormally distributed allow us to derive the following closed-form expressions for the distribution of the terminal wealth $V_T^\Psi$ and its parameters.
		\begin{prop}[Distribution of the terminal Shiryaev portfolio value]\label{prop_distribution_shiryaev} 
			The mean of  the terminal portfolio value $V_T^\Psi$ of the continuous-time Shiryaev strategy  is given by 
			\begin{align}
				\label{mean_theo_Shiryaev}
				\E\big[V_T^\Psi\big]& = s_0^1\big(g(2\mu T,4\sigma^2T^{2H}) -2 g(\mu T,\sigma^2 T^{2H}) +1\big),		
			\end{align}
			where $g(m,\Sigma)=\exp\{m+\Sigma/2\}$ is the mean of a lognormally distributed random variable with parameters $m$ and $\Sigma$. For the variance, we have $\Var{V_T^\Psi}= \E[(V_T^\Psi)^2]-(\E[V_T^\Psi])^2$ with the second-order moment 
			\begin{align}			
				\E\big[(V_T^\Psi)^2\big]& = (s_0^1)^2\big(g(4\mu T,16\sigma^2T^{2H}) -4 g(3\mu T,9\sigma^2 T^{2H}) \\
				& \qquad~~~~~ + 6g(2\mu T,4\sigma^2T^{2H}) -4 g(\mu T,\sigma^2 T^{2H}) +1 \big).
			\end{align}
			The cumulative distribution function of \,$V_T^\Psi$ is  
			\begin{align}						
				F_{V_T^\Psi} (v) = \P(V_T^\Psi\le v) &= \overline \Phi \Big(\frac{1}{\sigma T^H}\Big(\log\Big\{1+\sqrt{{v}/{s_0^1}}\Big\}-\mu T\Big)\Big) \\
				&                - \overline \Phi \Big(\frac{1}{\sigma T^H}\Big(\log\Big\{\Big(1-\sqrt{{v}/{s_0^1}}\,\Big)^+\Big\}-\mu T\Big)\Big), \quad v\ge 0,	
			\end{align}
			where $\overline \Phi$ denotes the cumulative distribution function of the standard normal distribution, $x^+=\max(x,0)$ for $x\in\R$, and 	 $\log 0$ is set to $-\infty$.
			
		\end{prop}	
		\begin{proof}
			See Appendix \ref{prop_distribution_shiryaev_proof}.
		\end{proof}
	}
	
	\paragraph{Salopek strategy}
	Another arbitrage strategy, dating back to \cite{Harrison1984} and applied in a fractional Black-Scholes market by \cite{Salopek1998}, trades $d\ge2$ risky assets and ignores the risk-free asset. It is defined for two real-valued constants $\alpha<\beta$ and can be summarized as $\stratc=(0,\stratc(\alpha,\beta))$ or, with some abuse of notation, $\stratc=\stratc(\alpha,\beta)$. The entries of $\stratc(\alpha,\beta)=(\Psi_t^1(\alpha,\beta),\ldots,\Psi_t^d(\alpha,\beta))$ are the risky asset shares at time $t\in[0,T] $. Specifically, for $i=1,\ldots,d$, we have 
	\begin{align}\label{eq:continuousSalopekStrategy}
		\Psi_{t}^i(\alpha,\beta)= \widehat\Psi_t^i(\beta)-\widehat\Psi_t^i(\alpha), \quad
		\text{where }\quad 
		\widehat{\Psi}_t^i(a)
		=\frac{1}{d}\left(\frac{S_t^i}{M_a(S_t)}\right)^{a-1}.
	\end{align}
	$M_a(x)$ denotes the $a$-order power mean of $x=(x^1,\ldots,x^d)\in\R_+^d$. It is given by
	\begin{equation}\label{eq:alphaorder}
		\begin{aligned}
			&M_a(x)=\Big(\frac{1}{d}\sum_{i=1}^{d}(x^i)^a\Big)^{1/a} & \text{for} & \quad a\neq 0,\\
			&M_0(x)=\sqrt[d]{x^1\cdot\ldots \cdot x^d} & \text{for} & \quad a = 0.
		\end{aligned}
	\end{equation}
	To provide an economic interpretation of this strategy, it is instructive to recall some properties of the involved family of power means \cite[see][]{Hardy1934}.
	\begin{bem}\label{rem:mean}With respect to the properties of the $a$-order power mean, we can list the following important special cases:
		\[\begin{array}{lll}
			M_1(x)& =(x^1+\ldots+x^d)/d & \qquad(\text{arithmetic mean})\\ [0.5ex]
			M_2(x)& =\sqrt{((x^1)^2+\ldots+(x^d)^2)/d} & \qquad(\text{quadratic mean})\\[0.5ex]
			M_{-1}(x)& =\big((1/x^1+\ldots+1/x^d)/d\big)^{-1} & \qquad(\text{harmonic mean})\\[0.5ex]
			M_0(x)&=\sqrt[d]{x^1\cdot\ldots \cdot x^d}=\lim_{a\to 0} M_a(x) & \qquad(\text{geometric mean})\\[0.5ex]
			M_\infty(x)&:=\lim_{a\to +\infty} M_a(x)=x_{max}=\max\{x^1,\ldots,x^d\} & \qquad(\text{maximum of  }x)\\[0.5ex]
			M_{-\infty}(x)&:=\lim_{a\to -\infty} M_a(x)=x_{min}=\;\min\{x^1,\ldots,x^d\} & \qquad(\text{minimum of }x)
			\\		
		\end{array}
		\]    
		Furthermore, the function $a\mapsto M_a(x)$ is increasing. For $a<b$, we have 
		\begin{align}
			\label{eq:mean_monoton}
			x_{min}\le M_a(x)\le M_b(x)\le x_{max} 
		\end{align}
		with equalities if and only if $x^1=\ldots=x^d=\overline x$, i.e., all entries of $x$ are identical. In this situation, $M_a(x)=\overline x$ holds for all $a\in\R$. 
	\end{bem}	
	
	The strategy $\stratc(\alpha,\beta)$ in \eqref{eq:continuousSalopekStrategy} is expressed as the difference between $\widehat\stratc(\beta)$ and $\widehat\Psi(\alpha)$. Because these two components can be considered as strategies themselves, we call $\widehat\Psi(a)$ an \textit{$a$-strategy} or \textit{$a$-portfolio}. Consequently, an investor can implement $\stratc(\alpha,\beta)$ by purchasing a $\beta$-portfolio and short-selling an $\alpha$-portfolio.
	
	Substituting the $\widehat\Psi(a)$ specified by \eqref{eq:continuousSalopekStrategy} into \eqref{eq:continuousValue} provides the portfolio value of an $a$-strategy, i.e., we obtain
	\begin{align}
		\label{eq:value_a_strat}
		V_t^{\widehat\stratc(a)}= M_a(S_t).
	\end{align}
	It also shows that $V_0^{\widehat\stratc(a)}= M_a(s_0)$, i.e., the initial investment is positive and equals the $a$-order power mean of initial asset prices $s_0$. An $a$-strategy investor enters long positions in all $d$ risky assets and chooses their numbers proportional to $(S_t^i)^{a-1}$. More specifically, for $a=1$, we have $\widehat\Psi_t^i=1/d$, i.e., an equally allocated investment, and the portfolio value is $V_t^{\widehat\stratc(1)}= M_1(S_t)$, i.e., the arithmetic mean of prices. For $a>1$ ($a<1$), the portfolio contains more (fewer) high-priced assets than low-priced assets. This feature becomes more pronounced with higher (lower) orders $a>1$ ($a<1$). In the limit for $a\to \infty$ ($a\to -\infty$), the investor only holds the asset with the highest (lowest) price. If there are $m\ge 1$ risky assets sharing this price, he orders $1/m$ each. 
	
	The strategy \eqref{eq:continuousSalopekStrategy} is an arbitrage strategy if we impose the following assumption on the financial market model.
	\begin{ass}
		\label{ass:AssetPrices}
		All price processes $(S_t^i)_{t\in[0,T]}$	of the risky assets $i=1,\ldots,d$ 
		start at time $t=0$ with identical initial prices $S_0^i=\widetilde s>0$.
	\end{ass}
	
	\begin{bem}
		Assumption \ref{ass:AssetPrices} serves mathematical simplification and will not be fulfilled in practice. However, this is not problematic because we can rescale the asset prices via $\widetilde S_t^i=\frac{\widetilde s}{s_0^i}S_t^i$ to $\widetilde S_0^i=\widetilde s$ and compute the arbitrage positions $\widetilde \Psi_t^i$ in the rescaled market. They are linked to the original market via  $\Psi_t^i=\frac{\widetilde s}{s_0^i}\widetilde \Psi_t^i$. Because $\Psi_t^i S_t^i=\widetilde \Psi_t^i \widetilde S_t^i$, \eqref{eq:continuousValue} delivers $V^{\stratc}_t=V^{\widetilde \stratc}_t$. That is, the portfolio value is not affected by the transformation.
	\end{bem}
	
	We now show that \eqref{eq:continuousSalopekStrategy} satisfies the three conditions in \eqref{eq:arbitrage} and is in fact an arbitrage strategy. As far as the self-financing property is concerned, it has been verified by \cite{Salopek1998}. According to \eqref{eq:value_a_strat}, for all $t\in[0,T]$, the portfolio value is given by 
	\begin{align}
		\label{eq:TermValueSalopek}
		V_t^{\stratc}=V_t^{\widehat\stratc(\beta)}-V_t^{\widehat \stratc(\alpha)}= M_\beta(S_t)- M_\alpha(S_t)\ge 0,
	\end{align}
	where we have used the assumption $\alpha<\beta$ and relation \eqref{eq:mean_monoton} stating that $M_a(\cdot)$ is increasing in $a$. This proves condition (ii) in \eqref{eq:arbitrage} {\neu with the lower bound $\arbbound=0$}. Condition (i) on zero initial investment follows from Assumption \ref{ass:AssetPrices} of identical initial asset prices. It yields $V_0^{\stratc}= M_\beta(s_0)- M_\alpha(s_0)=\widetilde s -\widetilde s= 0$.\footnote{Identical initial prices imply that, at time $t=0$, an investor formally buys $\widehat \Psi^i_0(\beta)=1/d$ shares of each asset and simultaneously sells $\widehat \Psi^i_0(\alpha)=1/d$ shares of each asset.} Finally, because we have assumed that the asset prices \eqref{eq:differentialequ} are driven by independent fBms, prices are uncorrelated and thus, at time $T$, almost surely not identical. This implies the strict inequality  $V_T^{\stratc}=M_\beta(S_T)- M_\alpha(S_T) > 0$ with probability one such that arbitrage condition (iii) is also satisfied.
	
	{\neu The result in \eqref{eq:TermValueSalopek} is the starting point for characterizing the distribution of $V_T^{\stratc}$. However, contrary to the terminal Shiryaev portfolio value covered by Proposition \ref{prop_distribution_shiryaev}, closed-form distribution expressions cannot be obtained in the Salopek case. Instead, the following numerical integration method allows arbitrarily precise approximation.}
	
	{\neu \begin{bem}[Distribution of the terminal Salopek portfolio value]\label{rem_distribution_Salopek} According to \eqref{eq:TermValueSalopek}, the terminal portfolio value $V_T^\Psi$ of the continuous-time Salopek strategy is given by 
			$V_T^{\stratc}= M_\beta(S_T)- M_\alpha(S_T)$.  From the price model \eqref{eq:BlackScholes} and Assumption \ref{ass:AssetPrices}, we deduce 
			\begin{align*}
				S_T^i =  \widetilde{s}\, \exp\{\mu^i T +\sigma^iB_T^{H^i}\} =  \widetilde{s}\, \exp\{\mu^i T +\sigma^i T^{H^i}X^i\},
			\end{align*}
			where the $X^i=B_T^{H^i}/T^{H^i}, i=1,\ldots,d$, are  independent standard normal random variables. This permits writing the vector of terminal prices as $S_T=p(X)$, where $X=(X^1,\ldots,X^d)^\top$ is a $d$-dimensional standard normal random vector with the joint probability density function $\phi_d(x)=\phi(x^1)\cdot \ldots\cdot \phi(x^d)$ and $\phi(u)=exp\{-u^2/2\}/\sqrt{2\pi}$. Further, the function   $p:\R^d\to \R^d$ is defined by 
			\[p(x)=(p^1(x^1),\ldots,p^d(x^d))^\top, \quad  p^i(u)=\widetilde{s}\, \exp\{\mu^i T +\sigma^i T^{H^i}u\},~i=1,\ldots,d,~~ u\in \R.\] 
			To receive the mean, variance and cumulative distribution function $F_{V_T^{\stratc}}(v)$ of $V_T^{\stratc}$, we need to compute expectations of functions $g(V_T^{\stratc})$ with some $g:\R\to\R $. For the mean and second-order moment, we set $g(u)=u$ and $g(u)=u^2$, respectively. As far as $F_{V_T^{\stratc}}(v)$ is concerned, we use $g(u)= \mathds{1}_{\{u\le v\}}(u)$ with indicator function $\mathds{1}$. These expectations can be expressed as $d$-fold integrals
			\begin{align*}
				\E\big[g(V_T^{\stratc})\big] & = \int_{\R^d} g\big(M_\beta(p(x))- M_\alpha(p(x)  \big) \phi_d(x) dx
			\end{align*}
			and evaluated by numerical integration. In the case of $d=2$ assets, which is the focus of our simulation study, we have double integrals of the form
			\begin{align*}
				\E\big[g(V_T^{\stratc})\big] & = \int_{-\infty}^\infty \int_{-\infty}^\infty g\big(M_\beta(p(x^1,x^2))- M_\alpha(p(x^1,x^2)  \big) \phi(x^1)\phi(x^2) dx^1\, dx ^2.
			\end{align*}
			\vfill
		\end{bem}
	}
	
	The monotonicity property \eqref{eq:mean_monoton} of the $a$-order power mean and the portfolio value expression \eqref{eq:TermValueSalopek} suggest to choose the largest possible $\beta$ and the smallest possible $\alpha$. Fusing the limits $\beta\to\infty$ and $\alpha\to-\infty$ into the following proposition shows that an arbitrage strategy with large $d$ can reduce to just buying the asset $i$ with the highest price and short-selling the asset $j$ with the lowest price.
	
	\begin{prop}\label{prop_Salopek}	
		Let $i,j\in \{1,...,d\}$, $t\in[0,T]$ and $\alpha<1\le \beta$ such that prices satisfy
		\begin{align}
			\label{eq:Salopek_monoton}
			S_t^i	> M_\beta(S_t)  > M_\alpha(S_t) > S_t^j. 
		\end{align}	
		Then, the strategy \eqref{eq:continuousSalopekStrategy} has the property   $\stratc_t^i(\alpha,\beta)>0>\stratc_t^j(\alpha,\beta)$. That is, the investor buys the high-priced $i$ and short-sells the low-priced $j$. This particularly holds when the prices of $i$ and $j$ represent the maximum and minimum over all $S_t^1,\ldots,S_t^d$.
	\end{prop}
	\vspace{-0.5cm}
	\begin{proof}
		See Appendix \ref{prop_Salopek_proof}.
	\end{proof}
	
	\subsection{Discrete-time arbitrage}\label{subsec:DisFracFinModel}
	We now replace the idealized continuous-time financial market model with permanent and frictionless asset transfers by a more realistic setup where trading takes place only at a finite number of fixed points in time and is subject to transaction costs.
	
	\paragraph{Discrete-time trading}
	In the discrete-time financial market model, prices are quoted at the times $0=t_0<t_1<\ldots <t_N=T$. A portfolio is created at time $t_0=0$, rebalanced at times $t_1,\ldots,t_{N-1}$ and liquidated at terminal time $t_N=T$. {\neu We focus on equidistant instants of time $t_n=n\Delta t, n=0,\ldots,N$, which divide the total trading horizon $[0,T]$ into $N$ trading periods of the same length $\Delta t=T/N$.}
	Thus, sampling the asset price processes \eqref{eq:BlackScholes} of the fractional Black-Scholes model  at $t_0,\ldots, t_N$ generates a sequence of risk-free asset prices  $\left(S_{t_n}^0\right)_{n=0,...,N}$ and $d$ sequences of  risky asset prices $\left(S_{t_n}^i\right)_{n=0,...,N}$ defined by
	\begin{align}\label{eq:FinancialModel}
		S_{t_n}^0=1,\quad S_{t_n}^i=s_0^i \exp\big\{\mu^i t_n+\sigma^i B_{t_n}^{H^i}\}, \quad \text{for} \quad n=0,\ldots,N \text{ and } i=1,\ldots,d,
	\end{align}
	with the same parameters as in Section~\ref{subsec:ContinuousModel}. 
	
	It has to be expected that the discretization of a self-financing continuous-time strategy $\Psi$, such as the Shiryaev and Salopek strategies of Section \ref{sec:ArbitrageStrategies}, and the existence of transaction costs affect the self-financing property. Even without transaction costs, rebalancing a portfolio according to a discretized self-financing continuous-time strategy requires the infusion of or allows the withdrawal of capital.\footnote{In Proposition \ref{Shiryaev_rebalancing}, we show that, for example, discretizing the Shiryaev strategy \eqref{eq:Shiryaev_stratc} almost surely leads to  a strictly positive capital requirement.} These rebalancing costs and the classic transaction costs may be incorporated by modifying the risk-free asset holdings $\Phi^0$ defined below. However, because we wish to explicitly quantify the impact of time discretization and transaction costs on  continuous-time arbitrage strategies, we extend our financial market model by an additional asset $d+1$ which we call transaction account. In the investment fund industry, such (cash) accounts are used to react flexibly to market events \cite[see][]{Nascimento2010,Simutin2014}. In our context, it allows us to express the aforementioned impact in monetary units.
	Similar to the risk-free asset price $S^0$, the price process of the new asset is a constant process $S^{d+1}=(S^{d+1}_{t_n})_{n=0,\ldots,N}$ with $S^{d+1}_{t_n}=1$. 
	
	We capture the trading activity of an investor by an initial capital amount $\Vini\in\R$ and the discrete-time $\mathbb{F}$-predictable process 
	\begin{align}
		\label{eq:strat_discrete}
		\stratd= (\stratd_n)_{n=1,\ldots,N+1}=(\Phi_n^0,\Phi_n^1,\ldots,\Phi_n^{d+1})_{n=1,\ldots,N+1}.
	\end{align}
	Here, $\Phi_n^0$ and $\Phi_n^{d+1}\in\mathbb{R}$ denote for $n=1,...,N$ the holdings in the risk-free asset and the transaction account, respectively, chosen at the beginning of the $n$-th trading period $[t_{n-1},t_n)$ and kept constant over that period. 
	Further,  $\Phi_n^i\in\mathbb{R}$ is the quantity of risky asset $i=1,\ldots,d$ held in the $n$-th trading period. The vector  $\Phi_{N+1}$ relates to the liquidation of the portfolio at time $t_N=T$.
	Overall, $\Phi$ is the \textit{discrete-time} \textit{portfolio} or \textit{trading strategy}.
	
	Discretizing the continuous-time strategy  $\stratc=(\Psi_t)_{t\in[0,T]}$ in \eqref{eq:strat_cont} leads to a piece-wise constant strategy where the investor period-wise sets and upholds $\stratc$. This means that, for  $i=0,\ldots,d$, we have $\Phi_n^i=\Psi_{t_{n-1}}^i, n=1,\ldots, N$, whereas liquidating the portfolio yields  $\Phi_{N+1}^i=0$. The positions $\Phi_n^{d+1}$ in the transaction account are specified in what follows.  
	
	\paragraph{Transaction costs} For purchasing, rebalancing and liquidating the portfolio, the investor has to pay transaction costs depending on the \textit{trading volume} of the risky assets. For a given strategy $\stratd$, at time $t$, this volume is defined by   
	\begin{align}
		\label{eq:trading_vol}
		\Gamma^{\stratd}_{t}=
		\left\{
		\begin{array}{{lll}}
			\sum_{i=1}^{d}\left|\Phi_{1}^i\right|S_{0}^i, &  t=t_0=0, & \text{(purchasing)}\\[1ex]
			\sum_{i=1}^{d}\left|\Phi_{n+1}^i-\Phi_{n}^i\right|S_{t_n}^i, &  t=t_1,\ldots, t_{N-1}, & \text{(rebalancing)}\\[1ex]
			\sum_{i=1}^{d}\left|\Phi_{N}^i\right|S_{T}^i, &  t=t_N=T. & \text{(liquidating)}\\
		\end{array}
		\right.
	\end{align}
	We specify transaction costs proportional to the trading volume. They are determined by the proportionality factor $p_1\ge 0$ (in \textit{percent}) if they exceed the minimum fee $p_2\ge 0$ (in \textit{monetary units}). Otherwise, $p_2$ is charged. We denote $p=(p_1,p_2)$ and define the \textit{transaction costs} for $  t=t_0,\ldots,t_N$ as 
	\begin{align}
		\label{eq:trans_cost}
		L_{t}^{\stratd} =l(\Gamma^{\stratd}_{t},p) \quad \text{with}  \quad l(y,p)=\max(p_1y,p_2)\one_{\{y>0\}}.
	\end{align}
	Note that no transaction costs are charged at time $t$ if the trading volume $\Gamma^\Phi_t$ is zero. The special case of a model without transaction costs is reflected by $p_1=p_2=0$.
	
	\paragraph{Liquidation} In the continuous-time model with frictionless trading, the terminal portfolio value $V_T^{\stratc}$ in \eqref{eq:continuousValue} is equal to the revenue from selling the portfolio. In the discrete-time case, liquidating the portfolio induces the transaction costs 
	$L^{\stratd}_{t_N}=l(\Gamma^{\stratd}_{t_N},p)$ of \eqref{eq:trans_cost} and \eqref{eq:trading_vol}.
	Thus, the \textit{net revenue} is 
	\begin{align}
		\label{eq:revenue_liq}
		R^{\stratd} =\sum_{i=0}^{d}\Phi_N^iS_{T}^i - L_{T}^{\stratd}.
	\end{align}
	
	\paragraph{Transaction account} As discussed above, we have augmented our model by an asset $d+1$ called \textit{transaction account}. It is used to finance rebalancing and transaction costs. Furthermore, it receives the net liquidation revenue at terminal time $t_N=T$. We now derive the holdings $\Phi^{d+1}_n, n=1,\ldots,N+1$, for this asset. To this end, the \textit{rebalancing costs} of a strategy $\stratd$ at time $t_n$ are denoted by $D_{t_n}^{\stratd}$ and defined as the value difference between the (risk-free and risky) asset holdings after and before trading:
	
	\begin{equation}\label{eq:Difference}
		\begin{aligned}
			D_{t_n}^{\stratd} =\sum\limits_{i=0}^{d} \Phi_{n+1}^i S_{t_n}^i -   \sum\limits_{i=0}^{d} \Phi_{n}^i S_{t_n}^i 
			&= \sum\limits_{i=0}^{d} \big(\Phi_{n+1}^i-\Phi_n^i\big)S_{t_n}^i\\ 
			&= \sum\limits_{i=0}^{d} \big(\Psi_{t_n}^i-\Psi_{t_{n-1}}^i\big)S_{t_n}^i,
			~ n=1,\ldots,N-1.
		\end{aligned}
	\end{equation}
	Here, the last line follows from the sampling property $\Phi_n^i=\Psi_{t_{n-1}}^i, i=1,\ldots,d$.
	Also note that the rebalancing costs at time $t_0=0$ and $t_N=T$ are zero.
	
	Aggregating the rebalancing and transaction costs as well as the net liquidation revenue, the holdings in the transaction account can be stated recursively by
	\begin{align}
		\nonumber
		{\Phi}_{1}^{d+1}&=-L_{0}^{\stratd},&& (\text{purchasing})\\
		\label{eq:rebal_strat}
		{\Phi}_{n+1}^{d+1}&={\Phi}_{n}^{d+1} - D_{t_n}^{\stratd}-L_{t_n}^{\stratd},\quad n=1,\ldots,N-1,
		&& (\text{rebalancing})\\
		\nonumber
		{\Phi}_{N+1}^{d+1}&={\Phi}_{N}^{d+1} +R^{\stratd},
		&& (\text{liquidating})
	\end{align}
	where $L_{t}^{\stratd}$, $R^{\stratd}$ and $D_{t}^{\stratd}$ are obtained according to \eqref{eq:trans_cost}, \eqref{eq:revenue_liq} and \eqref{eq:Difference}, respectively.
	
	\paragraph{Portfolio value}
	At time $t_n$, the {value} $V_{t_n}^{\stratd}$ of the portfolio $\stratd$ is
	\begin{align}
		V_{t_n}^{\stratd} = \sum_{i=0}^{d+1} \Phi_{n+1}^i S_{t_n}^i, \quad n=0,\ldots,N.\label{eq:discreteValue}
	\end{align}  
	$V^\stratd=(V_{t_n}^{\stratd})_{n=0,\ldots,N}$ is the \textit{discrete-time value process}. While, at time $t=0$, the continuous-time model yields   $V_{t_0}^{\stratc}=\Vini$, the discrete-time case delivers $V_{0}^{\stratd}=\Vini-L_{0}^{\stratd}$. That is, the portfolio value equals the initial capital minus the transaction costs for purchasing the portfolio. If $\stratd$ results from discretizing a continuous-time arbitrage strategy $\stratc$ with $V_0^\stratc=v=0$, the discrete-time value process starts with $V_{0}^{\stratd}=-L_{0}^{\stratd}\le 0$. This term is strictly negative if the initial trading volume $\Gamma_0^\stratd$ and at least one of the two transaction cost parameters $p_1$ and $p_2$ are positive. For the terminal trading time $t_N=T$, substituting $\Phi^0_{N+1}=\ldots =\Phi^d_{N+1}=0$ into  \eqref{eq:discreteValue} and applying \eqref{eq:rebal_strat} provides  $V_{T}^{\stratd}=\Phi_{N+1}^{d+1}S_{t_N}^{d+1} ={\Phi}_{N}^{d+1} +R^{\stratd}$. Hence, the terminal portfolio value equals the net liquidation revenue $R^{\stratd}$ minus the cumulated rebalancing and transaction costs  ${\Phi}_{N}^{d+1}$ for trading at times $t_0,\ldots,t_{N-1}$. 
	
	\begin{bem}
		\label{rem_scaling_discrete}
		For a continuous-time arbitrage strategy $\stratc$, we know from Remark \ref{rem_scaling} that scaling the strategy by some factor $\scalefac>0$ preserves the arbitrage property. $\widetilde \stratc=\scalefac \stratc$ is also an arbitrage strategy. For the value process, we have $V^{\scalefac\stratc}=\scalefac V^{\stratc} $.
		A time discretization of $\stratc$ and transaction costs generally destroy the arbitrage property. However, inspecting the construction of the discretized strategy $\stratd$ reveals that we preserve the scaling property of the discrete-time value process $V^{\scalefac\stratd}=\scalefac V^{\stratd} $ as long as the transaction costs are defined with a floor $p_2=0$, i.e., only proportional transaction costs $L_t^\stratd=p_1 \Gamma_t^\stratd$ are charged.  
	\end{bem}
	
	Finally, it is noteworthy that the discrete-time value process $V^\stratd$ satisfies a generalized self-financing condition
	\begin{align*}
		V_{t_n-}^{\stratd} - L_{t_n}^\stratd=  V_{t_n}^{\stratd}, \quad n=0,\ldots, N-1,
	\end{align*}
	where $V_{t_n-}^{\stratd}=\sum_{i=0}^{d+1} \Phi_{n}^i S_{t_n}^i$ is the portfolio value before rebalancing at $t_n, n=1,\ldots, N-1$, and $V_{0-}^{\stratd}=\Vini$. This condition formalizes the property that the value after rebalancing equals the value before rebalancing minus the transaction costs of the corresponding trade.
	
	{\neu
		\paragraph{Portfolio value differences}
		The continuous and discrete portfolio values \eqref{eq:continuousValue} and \eqref{eq:discreteValue}, respectively, $\Phi_n^i=\Psi_{t_{n-1}}^i, n=1,\ldots, N,$ and $S_t^{d+1}=1$ allow us to express the performance of the discretized strategy $\stratd$ relative to $\stratc$ in terms of transaction account holdings:
		\begin{align}
			\label{eq:value_cont_discrete}
			V_t^\stratd- V_t^\stratc =\Phi_t^{d+1}  \quad \text{for } \quad t=t_0,\ldots, t_{N-1}.
		\end{align}
		While \eqref{eq:value_cont_discrete} only holds before liquidation, the following lemma presents a more general relationship for all times $t$, i.e., including the liquidation date $t=t_N=T$.
		
		\begin{lem}\label{lemma}
			For $t=t_0,\ldots, t_{N}$, the difference between the portfolio values $V_t^\stratc$ and $V_t^\stratd$ can be fomulated as
			\begin{align} 
				\label{def_cum_rebalancing}
				V_t^\stratc- V_t^\stratd &= \rebalcost_t +\transactcost_t, \quad \text{with} \quad 
				\rebalcost_t = \sum_{t_n\le t} D^\stratd_{t_n}\quad \text{and} \quad  \transactcost_t = \sum_{t_n\le t} L^\stratd_{t_n},
			\end{align}
			where the summation terms $\rebalcost_t $ and $\transactcost_t$ denote the cumulated rebalancing and  transaction costs in the period $[0,t]$, respectively.
		\end{lem}
		\vspace{-0.5cm}
		\begin{proof}
			See Appendix \ref{lemma_proof}.
		\end{proof}
		
		\noindent In a situation without transaction costs, the decomposition in Lemma \ref{lemma} implies that the loss in terminal portfolio value, i.e., the \textit{discretization error}, going along with the discrete-time transfer of the continuous-time strategy $\stratc$ is
		\begin{align}
			\label{error_value}
			\rebalcosts = V_{T}^{\stratc} - V_{T}^{\stratd}.
		\end{align}
		In Propositions \ref{prop_conv_shiryaev} and \ref{prop_conv_salopek} below, we study the asymptotic behavior of this quantity for an increasing trading frequency, i.e., for $N\to \infty$ such that  $\Delta t \to 0$.	
	}
	
	\section{Simulation study}\label{sec:Analysis}   
	\subsection{Parameters {\neu and data generation}}\label{subsec:ModelParameter} 
	
	To provide a full-scale analysis of our two arbitrage strategies, we conduct a Monte Carlo study {\neu and a complementary mathematical analysis of discretization error} based on the model and trading parameters of Table \ref{Tab:paramters}. This table captures our \textit{basis setting} which will be successively modified and relaxed as we proceed.
	
	\begin{table}[h!]
		\footnotesize
		\begin{center}
			\begin{tabular}{llll}
				\toprule
				Risky assets & Number of assets & $d$& $1$ (Shiryaev), $2$ (Salopek)\\
				& Drift 	& $\mu^i$	& $0.05$\\
				& Volatility 			& $\sigma^i$		& $ 0.1$ \\
				& 	Hurst coefficient	& $H^i$ & $0.6$\\ 
				& Initial value  & $s_0^i=\widetilde s$ & $100$\\
				\midrule
				Trading	
				& 		Trading horizon				& $T$			& $1$  \\
				&		Trading periods	&	$N$& $250$ \\
				&		Trading dates				& $t_n$	 & $n\Delta t = {n}/{N},\quad   n=0,\ldots,N$\\	
				&		Transaction costs & $p=(p_1,p_2)$  & $(0,0)$\\
				&		Scaling factor &$\scalefac$  & $100$\\
				&       Salopek specification & $(\alpha,\beta)$ &$(-30,30)$\\
				\bottomrule
			\end{tabular}
		\end{center}
		\scriptsize
		This table summarizes the model parameters of the risky assets $i=1,2$ (see Sections \ref{subsec:ContinuousModel} and \ref{sec:ArbitrageStrategies}) and the discrete-time trading parameters (see Sections \ref{sec:ArbitrageStrategies} and \ref{subsec:DisFracFinModel}) we use in our simulation basis setting.
		\caption{\label{Tab:paramters} Basis setting}
	\end{table}
	
	Guided by the empirical literature \cite[see][]{Willinger1999,Bessembinder2018}, we start by specifying suitable drifts $\mu^i$, volatilities $\sigma^i$ and Hurst parameters $H^i$ for the $d$ risky assets of the Shiryaev and Salopek strategies. We restrict the latter to $d = 2$ assets and assume that they have identical parameters. We also set the initial prices to $\widetilde s = 100$.
	
	We then consider an investor with a $T=1$ year investment horizon subdivided into $N = 250$ trading days \cite[see][]{Hendricks2005}. This investor is assumed to follow the discretization of Section \ref{subsec:DisFracFinModel} to trade the strategies of Section \ref{sec:ArbitrageStrategies} at a daily frequency. All transaction costs $p$ are zero. To capture the performance of this investor, we simulate 100,000 asset price scenarios and scenario-wise document $V_T^{\stratd}$, i.e., the portfolio value after liquidation.\footnote{This number of simulation repetitions ensures stable results \cite[see][]{Schuhmacher2011}.} Consequently, for each strategy (and parameter setting), our Monte Carlo study delivers a distribution of $V_T^{\stratd}$ values which will undergo detailed analysis.
	
	{\neu To generate a path of a discrete-time fBm many algorithms have been proposed in the literature \cite[see][]{Kijiama13}. They can be grouped into exact and approximate methods. While the first category fully preserves the key properties of a fBm at the cost of higher computation time, the second group sacrifices some accuracy to improve time efficiency. Practitioners are often interested in quick results and therefore tend to favor spectral simulation techniques of the approximate class \cite[see][]{Dieker2003}. We opt for an exact approach, i.e., the circulant embedding method of \cite{Wood1994}, because we wish to rule out any influcence of approximation error on our evaluation of strategy performance.\footnote{{\neu Independent of \cite{Wood1994}, the study of \cite{Dietrich1997} has developed the same simulation technique. \cite{Stein2002} discusses non-stationary surface extensions. In an earlier draft, we used the closely related approximate method of \cite{Yin1996}. It delivered slightly different performance metrics but allowed the same overall conclusions.}} This method competes quite well with the speed of approximate methods \cite[see][]{Coeurjolly2000} and, given the fact that simulation source code is readily available, its implementation is straightforward \cite[see][]{Kroese2015}.\footnote{{\neu We follow the starting value recommendation of \cite{Danudirdjo2011}. Due to the Fourier transform nature of the method, a single simulation run always generates two independent fBm realizations \cite[see][]{Dietrich1997}.}}}
	
	\subsection{Discretized Shiryaev strategy}
	\label{sec:Analysis_Shiryaev} 
	
	\subsubsection{{\neu General strategy behavior}}\label{subsec:BasicSetting} 
	
	We start with the Shiryaev strategy which trades the risk-free asset and one risky asset. For this strategy and our basis setting of Table \ref{Tab:paramters}, Figure~\ref{Fig:Shiryaev1} presents a typical realization of our simulations. Panel (a) plots the daily prices $S^0_{t_n}=1$ and $S^1_{t_n}$ of the risk-free and the risky asset, respectively. Panel (b) describes the daily strategy $\stratd$ and the associated rebalancing costs $D_{t_n}^{\stratd}$. 
	For visual convenience, we scale the number of risky assets by the initial risky asset price $S_0^1=\widetilde s$ and show the \textit{negative} costs $-D_{t_n}^{\stratd}$. Because of $p=(0,0)$ and \eqref{eq:rebal_strat}, the holdings $\Phi_n^2$ in the transaction account result from cumulating the rebalancing costs, i.e., $\Phi_1^2=0, \Phi_{n+1}^2=\Phi_n^2-D_{t_n}^{\stratd}, n=1,\ldots,N-1$. Finally, Panel (c) illustrates the value process $V^\stratc$ of the continuous-time strategy $\stratc$, the value process  $V^\stratd$ of the discretized strategy $\stratd$ and the difference between them. As shown in \eqref{eq:value_cont_discrete}, this difference equals the holdings $\Phi^2_{t_n}$ in the transaction account for $t_n=t_0,\ldots,t_{N-1}$. 
	
	\afterpage{
		\begin{figure}[ht!]
			\begin{center}
				\includegraphics[trim= 10mm 10mm 0mm 10mm, clip, width=12cm]{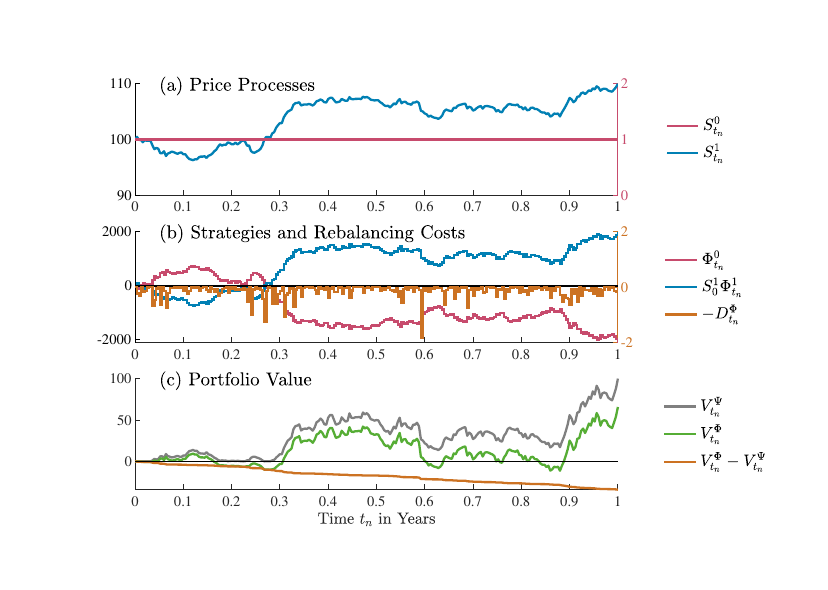}
			\end{center} 
			\scriptsize
			For the discretized Shiryaev strategy and our basis setting of Table \ref{Tab:paramters}, this figure plots a typical simulation result. Panel (a) shows the realized prices $S^0_{t_n}$ and $S^1_{t_n}$ of the risk-free and the risky asset, respectively. Panel (b) illustrates the strategy holdings $\Phi^0_{t_n}$ and $\Phi^1_{t_n}$ for these assets where the latter has been multiplied by $S_0^1$. Furthermore, it contains the negative rebalancing costs $-D_{t_n}^{\Phi}$. Finally, Panel (c) reports the portfolio value $V_{t_n}^{\stratd}$ of the strategy. It is supplemented by the value process $V_{t_n}^{\stratc}$ of continuous-time trading  and the difference between both portfolio values which, except for terminal time $t_N = T= 1$, is equal to the holdings $\Phi^2_{t_n}$ in the transaction account. 
			\caption{Exemplary realization of the discretized Shiryaev strategy}\label{Fig:Shiryaev1}
		\end{figure}
	}
	
	According to \eqref{eq:Shiryaev_stratc}, a Shiryaev-type investor enters a long (short) position in the risky asset whenever the risky asset price exceeds (falls below) the initial risky asset price. The opposite applies to the risk-free asset. This can be seen in Panels (a) and (b). We also observe that the rebalancing costs $D_{t_n}^{\stratd}$ are small and always positive; this is validated in Proposition \ref{Shiryaev_rebalancing}. Thus, each rebalancing activity requires new capital and increases the absolute value of the negative holdings $\Phi^2$ in the transaction account. In Panel (c), this leads to a growing difference between the portfolio values of continuous and discrete trading.
	
	\begin{prop}
		\label{Shiryaev_rebalancing}
		For the rebalancing costs \eqref{eq:Difference} of the discretized Shiryaev strategy \eqref{eq:Shiryaev_stratc} and $n=1,\ldots, N-1$, it holds almost surely that  		
		\begin{align}
			\label{rebalance_shiryav}
			D_{t_n}^{\stratd}	\neu =\frac{(S_{t_n}^1-S_{t_{n-1}}^1)^2}{s_0^1}>0.
		\end{align}
	\end{prop}
	\begin{proof}		
		See Appendix \ref{Shiryaev_rebalancing_proof}.
	\end{proof}
	
	We know from \eqref{eq:TermValueShiraev} that the portfolio value of the continuous-time arbitrage strategy  is $V^\Psi_t=(S_t^1-s_0^1)^2/s_0^1 \ge 0$. It rises with the distance between the risky asset price $S_t^1$ and the initial risky asset price $s_0^1$. In other words, the strategy benefits from prices rising above $s_0^1$ and from prices falling below $s_0^1$. As indicated in Remarks \ref{rem_scaling} and \ref{rem_scaling_discrete}, scaling the continuous-time strategy $\stratc$ by some factor $\scalefac>0$ preserves the arbitrage property and leads to a scaling of the value processes $V^\stratc$ and $V^\stratd$  by the same factor. Looking at the terminal value $V_T^{\stratd}\approx65$, we can deduce that raising the basis scaling factor of  $\scalefac=10^2$ to say $\scalefac=10^5$ increases the terminal value to roughly $65,000$. Hence, the absolute size of the portfolio value is not relevant for evaluating the performance of the strategy.
	
	\subsubsection{Impact of {\neu time discretization and} transaction costs}\label{subsec:transaction_Shiryaev}
	After examining a single simulation scenario in the parameter basis setting, we now turn to the results of 100,000 scenarios and additionally introduce transaction costs. 
	
	Panel (a) of Figure \ref{Fig:Shiryaev2} visualizes the simulated distribution of the terminal portfolio value $V_T^{\stratc}$ for continuous-time trading and the corresponding  $V_T^{\stratd}$ distributions for the discrete-time case with three different transaction cost variants. $p = (0,0)$ resembles no transaction costs. $p=(0.1,0)$ considers only proportional costs, whereas $p=(0.1,0.5)$ additionally includes a minimum fee. Recall that the proportional values are expressed in percent. The chosen cost magnitudes are guided by what are currently very low commissions and brokerage fees \cite[see][]{Auer2015liquid}. Table \ref{Tab:Shiryaev_statistics} provides summary statistics for the distributions and concisely evaluates the performance of the trading strategy. Here, we are particularly interested in the mean terminal value and the loss probability because they capture $\E[V_T^{\stratd}]$ and $\mathbb{P}(V_T^{\stratd}<0)$, respectively.
	
	\afterpage{
		\begin{figure}[ht!]
			\begin{center}
				\includegraphics[trim= 0mm 46mm 0mm 10mm, clip,width=12cm]{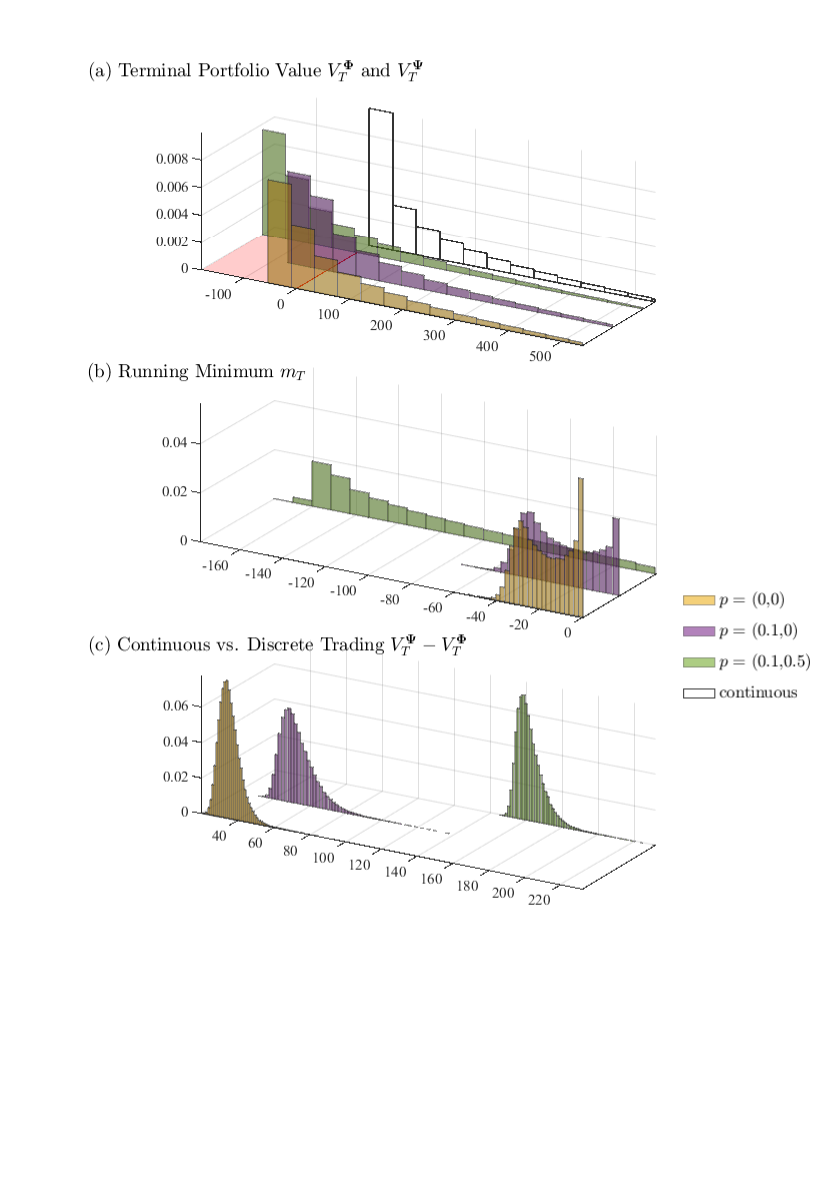}
			\end{center} 
			\scriptsize
			For 100,000 simulated scenarios of the discretized Shiryaev strategy, the basis setting of Table \ref{Tab:paramters} and a range of transaction cost values, this figure presents various portfolio value distributions. Panel (a) shows the distributions of the terminal value $V_T^{\stratd}$ of discrete-time trading with transaction costs $p = (p_1,p_2)$ where $p_1$ reflects proportional costs (in percent) and $p_2$ is a minimum fee (in monetary units). The loss region with negative terminal values is highlighted by a red floor. The distribution of the terminal value $V_T^{\stratc}$ of continuous-time trading is also included. Panel (b) contains the distributions of the running minimum $m_T$ of the discrete value processes, i.e., the worst-case portfolio values in the investment horizon. Finally, the distributions in Panel (c) refer to the terminal difference $V_T^\stratc- V_T^\stratd$ between continuous-time and discrete-time trading. For better visibility, the x-axis in Panel (a) is cut off after the 95\% quantile of $p=(0,0)$.
			\caption{Shiryaev portfolio value distributions for different transaction costs}\label{Fig:Shiryaev2}
		\end{figure}
		\vspace{-0.25cm}
		\begin{table}[ht!]
			\footnotesize
			\begin{center}
				\begin{tabular}[t]{lr|dd|ddddd|e}
					\toprule
					Strat.& Transact.& \mc{Mean} &  \mcl{Stand.} &  & \multicolumn{3}{c}{ Quantiles}  &  & \mc{Loss}  \\   	
					& costs $p$& & \mcl{dev.}    & \mc{Min} & \mc{5\%} & \mc{Median}   & \mc{95\%} & \mcl{Max} & \mc{prob.} \\
					\midrule
					$\stratc$ & none & 144.7 & 226.2 & 0.0 & 0.5 & 59.4 & 581.3 & 4,477.2 & 0.00 \\
					&  &  (144.2) &  (222.9) &  (0.0) &  (0.5) &  (59.6) &  (575.7) &  (\infty) &  (0.00) \\
					\midrule
					$\stratd$ & $(0,0)$ & 109.4 & 222.8 & -53.5 & -32.9 & 25.7 & 537.9 & 4,392.8 & 0.39 \\
					& $(0.1,0)$ &91.9 & 219.7 & -73.7 & -48.1 & 9.2 & 514.2 & 4,339.9 & 0.46 \\
					& $(0.1,0.5)$ &-17.3 & 221.0 & -178.5 & -157.9 & -100.4 & 407.8 & 4,245.5 & 0.73 \\
					\bottomrule
				\end{tabular}
			\end{center}
			\scriptsize 
			This table reports some descriptive statistics for the simulated terminal portfolio value distributions in Panel (a) of Figure \ref{Fig:Shiryaev2}. Besides the mean and standard deviation, we compute the minima and maxima as well as selected quantiles. Furthermore, we present the simulated loss probability, i.e., the proportion of negative terminal portfolio values. {\neu The numbers in parentheses for the continuous-time portfolio represent theoretical values obtained via Proposition \ref{prop_distribution_shiryaev}.}
			\caption{\label{Tab:Shiryaev_statistics} Shiryaev portfolio value statistics for different transaction costs}
		\end{table}
	}
	
	We observe that the Shiryaev strategy $\stratc$ is an arbitrage strategy with positive terminal values $V_T^{\stratc}$ in all scenarios. {\neu The distribution of $V_T^{\stratc}$ is characterized by positive skewness, a wide right tail and a mean of $144.7$. This simulated value is very close to its theoretical counterpart of $144.2$ which can be obtained via Proposition \ref{prop_distribution_shiryaev}.\footnote{\label{fn:ci}{\neu The asymptotic ($1-\alpha$) confidence interval for the mean $m$ is $\hat{m} \pm \frac{\hat{s}}{\sqrt{n}} z_{1-\alpha/2}$. For an empirical mean and standard deviation of $\hat{m} = 144.7$ and $\hat{s}=226.2$, respectively, a sample size of $n = 100,000$, and $\alpha = 5\%$, we have $(143.3, 146.1)$. The theoretical value of $m=144.2$ is covered by this range.}}} In comparison, discretizing the strategy yields a terminal value $V_T^{\stratd}$ distribution of similar shape but shifted towards smaller values and partially into negative territory. Without transaction costs, the range of observed terminal values is $[-53.5;4,392.8]$. This means that the maximum gain is significantly higher than the maximum loss. With a value of $109.4$, the mean of $V_T^{\stratd}$ is more than double the maximum loss, covers about 75\% of the mean of $V_T^{\stratc}$ and can be earned by accepting a loss probability of only 39\%.
	
	{\neu Proposition \ref{prop_conv_shiryaev} presents an asymptotic expansion of the expected cumulated Shiryaev rebalancing costs $\rebalcosts$ (with zero transaction costs) for an increasing trading frequency, i.e., $N\to\infty$ such that $\Delta t\to 0$. Because \eqref{error_value} implies $  V_{T}^{\stratd}	 = V_{T}^{\stratc}  - \rebalcosts$, it can be used to formally proxy $\E[V_T^{\stratd}]$ instead of depicting it via simulation. Specifically, the leading term $C\Delta t^{2H-1}$ of the expansion, where $C$ and $2H-1$ are the rate and order of convergence, respectively, delivers the approximation $109.0$ and proves to be quite accurate (see also Tables \ref{Tab:Shiryaev_mu} to \ref{Tab:Shiryaev_H} and, in particular, Table \ref{Tab:Shiryaev_freq}).}\footnote{{\neu We recognize the order $2H-1$ from the convergence to zero of the fBm's quadratic variation restricted to the time grid \cite[see][]{Lin1995}. The quality statement can be backed up similar to Footnote \ref{fn:ci}.}}
	
	{\neu
		\begin{prop}[Expected cumulated Shiryaev rebalancing costs]\label{prop_conv_shiryaev}
			For a discretization with step size $\Delta t=T/N, ~N\in \N$, and $\Delta t\to 0$, the expectation of the cumulated rebalancing costs of the Shiryaev strategy $\rebalcosts=\rebalcosts(\Delta t)=\sum_{n=1}^{N} D_{t_n}^{\stratd}$ given in \eqref{def_cum_rebalancing} satisfies 
			\begin{align}
				\E\big[\rebalcosts(\Delta t) \big]& = C\, \Delta t^{2H-1} + o(\Delta t^{2H-1}) \quad \text{with }~~ C=\sigma^2 s_0^1 \int_0^T \exp\big\{2\mu \,t+2\sigma^2 t^{2H}\big\}\,dt.
			\end{align}		
		\end{prop}
		\begin{proof}
			See Appendix \ref{prop_conv_shiryaev_proof}.  
		\end{proof}
	}
	
	In line with intuition, transaction costs shift the portfolio value distribution even further such that losses become higher and more likely. For $p=(0.1,0)$, the mean of $V_T^\stratd$ falls to $91.9$ but remains positive. Simultaneously, the loss probability rises to $46\%$ but can still be considered reasonably low \cite[see][]{Hogan2004}. In contrast, $p=(0.1,0.5)$ generates a negative mean of $-17.3$ and a loss probability of $73\%$. The reason for this drastic impact of the minimum fee is that our basis setting is of low monetary scale such that the daily trading volumes are small and the minimum fee applies frequently. For $N=250$ trades, this often results in a total fee of $250\times p_2=125$ offsetting gains and causing high losses. Overall, while proportional transaction costs only slightly reduce the performance of the discretized strategy, minimum fees can render it unattractive for small-scale investors. {\neu However, medium-scale investors with a higher scaling factor $\scalefac$ and thus higher trading volumes do not suffer from this kind of problem.}\footnote{{\neu In contrast to large-scale traders, medium-scale traders are often considered to have insignificant market impact \cite[see][]{Bouchaud2022}.}}
	
	Panel (b) of Figure \ref{Fig:Shiryaev2} conducts a worst-case analysis similar to drawdown calculations in active risk management \cite[see][]{Schuhmacher2011}. For $t\in [0,T]$, we define the running minimum process associated with the discrete-time value process $V^{\stratd}$ as
	\begin{align} 
		m_t:=\min_{t_n\le t}\, V_{t_n}^{\stratd}. 
		\label{eq:runningMinimum}
	\end{align}
	With $t=T$, we obtain $m_T$ 
	representing the least 
	favorable portfolio value in the investment horizon $[0,T]$. 
	The simulated distributions of $m_T$ show that its upper bound is zero because, across our transaction cost variants, we have $V_0^{ \stratd}\le0$. The smallest values of $m_T$ are close to the minima of the terminal values $V_T^{\stratd}$ in Panel (a). However, a frequency comparison reveals that the vast majority of worst-case events do not cluster at time $T$.
	
	Finally, Panel (c) shows the simulated distributions of the difference $V_T^\stratc-V_T^\stratd$  between the terminal portfolio values of continuous-time trading and its discrete-time counterparts. Discrepancies obviously rise with $p$. More demanding transaction cost variants require higher capital infusions, i.e., a more intensive usage of the transaction account.
	
	\subsubsection{Impact of asset model parameters}
	To implement the Shiryaev strategy, investors have to select a suitable risky asset. To support this choice, we study the impact of the asset parameters $(\mu,\sigma,H)$ on the performance of the strategy. This primarily involves constructing figures and tables in the style of Figure \ref{Fig:Shiryaev2} and Table \ref{Tab:Shiryaev_statistics}. However, instead of $p$, we now vary $(\mu,\sigma,H)$.
	
	\paragraph{Drift \boldmath$\mu$}
	Figure \ref{Fig:Shiryaev3} and Table \ref{Tab:Shiryaev_mu} start by considering the alternative drift parameters $\mu \in \{0,\pm0.1,\pm0.2\}$. Assets with higher absolute mean returns $\mu$ shift, flatten and extend the distribution of $V_T^\stratd$ towards larger terminal values because, {\neu \textit{ceteris paribus}}, they fuel the strategy with more extreme prices. This effect is stronger for positive than for negative $\mu$ because upward movements are unbounded, whereas downward movements have a floor at a price level of zero.\footnote{A similar rationale explains differences in the prices of at-the-money call and put options with identical underlying, strike and maturity \cite[see][]{Black1973}.} For example, starting from $69.2$ for $\mu=0$, the mean terminal value is $209.1$ for $\mu=0.1$ but only $136.1$ for $\mu=-0.1$. In contrast, the loss probabilities are quite symmetric in $\mu$. From 43\% for $\mu = 0$, they fall to about $28\%$ for $|\mu|=0.1$ and to $7\%$ for $|\mu|=0.2$. A similar feature can be observed for the running minimum $m_T$. For $\mu=0$, its distribution is almost uniform. For rising $|\mu|$, peaks near zero become more pronounced and excursions of the portfolio value significantly below zero less likely. $V_T^\stratc -V_T^\stratd$ is not symmetric in $\mu$. Instead, the distribution support increases with $\mu$ towards larger values. Hence, a higher $\mu$ induces more rebalancing costs in discrete-time trading.
	
	\afterpage{
		\begin{figure}[ht!]		
			\begin{center}
				\centering		
				\includegraphics[trim= 0mm 46mm 0mm 10mm, clip, width=12cm]{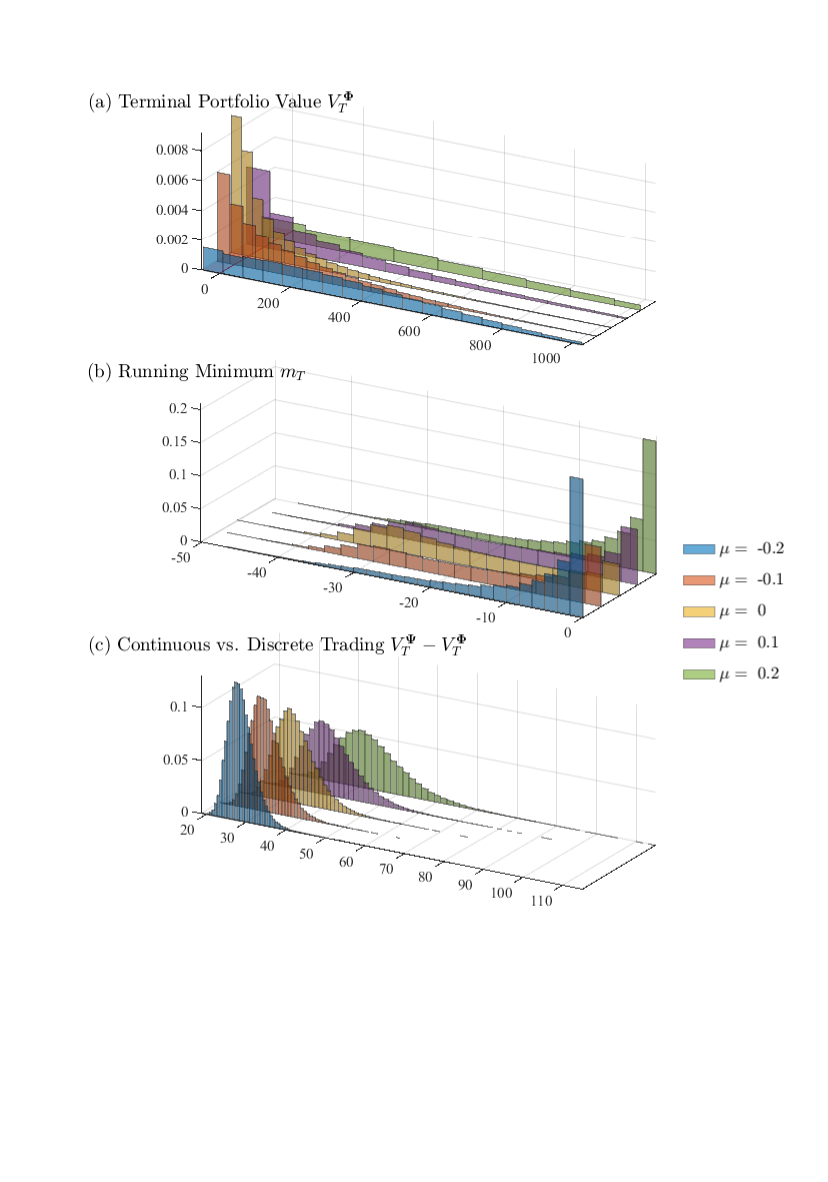}
			\end{center}		
			\scriptsize
			Similar to the Shiryaev strategy Figure \ref{Fig:Shiryaev2}, but for varying drift values $\mu$, this figure presents the simulated distributions of (a) the terminal portfolio value $V_T^{\stratd}$, (b) the running value process minimum $m_T$ and (c) the difference $V_T^\stratc- V_T^\stratd$ between the terminal portfolio values of continuous-time and discrete-time trading. For better visibility, the x-axis in Panel (a) ends at the 80\% quantile of  $\mu=0.2$.
			\caption{Shiryaev portfolio value distributions for different drifts}\label{Fig:Shiryaev3}
		\end{figure}
		
		\begin{table}[h!]
			\footnotesize
			\begin{center}
				\begin{tabular}{r|rrrrrrr}
					\toprule
					$\mu$& Mean & Median & Stand. dev.  & Min & Max & Loss prob. \\\midrule
					$-$0.2 & 355.0 \neu  (354.4) & 301.5 & 291.6 & $-$46.5 & 2,098.7 & 0.07 \\
					$-$0.1 & 136.1 \neu  (134.9) & 71.5 & 187.4 & $-$48.2 & 1,842.4 & 0.27 \\
					0.0 & 69.2 \neu $\;$ (68.3)& 12.8 & 153.2 & $-$50.1 & 3,377.4 & 0.43 \\
					0.1 & 209.1 \neu  (209.5)& 82.7 & 334.4 & $-$47.6 & 5,602.4 & 0.28 \\
					0.2 & 625.0 \neu  (628.0)& 446.1 & 641.8 & $-$46.7 & 8,706.9 & 0.07 \\
					\bottomrule
				\end{tabular}
			\end{center}
			\scriptsize
			Similar to Table \ref{Tab:Shiryaev_statistics}, this table reports some descriptive statistics for the simulated terminal portfolio value distributions in Panel (a) of Figure \ref{Fig:Shiryaev3}. {\neu The mean values in parentheses are approximations based on Proposition \ref{prop_conv_shiryaev}.}
			\caption{\label{Tab:Shiryaev_mu} Shiryaev portfolio value statistics for different drifts}
		\end{table}
	}
	
	\paragraph{Volatility \boldmath$\sigma$}
	Figure~\ref{Fig:Shiryaev4} and Table \ref{Tab:Shiryaev_sigma} present our sensitivity results for the volatilities $\sigma \in \{0.05,0.10,0.15\}$. Increasing volatility goes along with a greater variability of terminal portfolio values $V_T^\stratd$. Large gains and large losses become more likely. This is also evident in the stretching distributions of the running minimum $m_T$. The mean terminal values increase with $\sigma$ and so do the loss probabilities. For example, $\sigma=0.05$ delivers $46.6$ and $27\%$, whereas $\sigma=0.15$ yields $218.5$ and 41\%. Thus, if investors are prepared to bear more risk in exchange for higher reward, they have an incentive to opt for volatile assets \cite[see][]{Frazzini2014}. However, it must be noted that the rebalancing costs of discrete trading, which are resembled by $V_T^\stratc -V_T^\stratd$, substantially increase with $\sigma$.
	
	\afterpage{
		\begin{figure}[ht!]
			\begin{center}
				\includegraphics[trim= 0mm 46mm 0mm 10mm, clip, width=12cm]{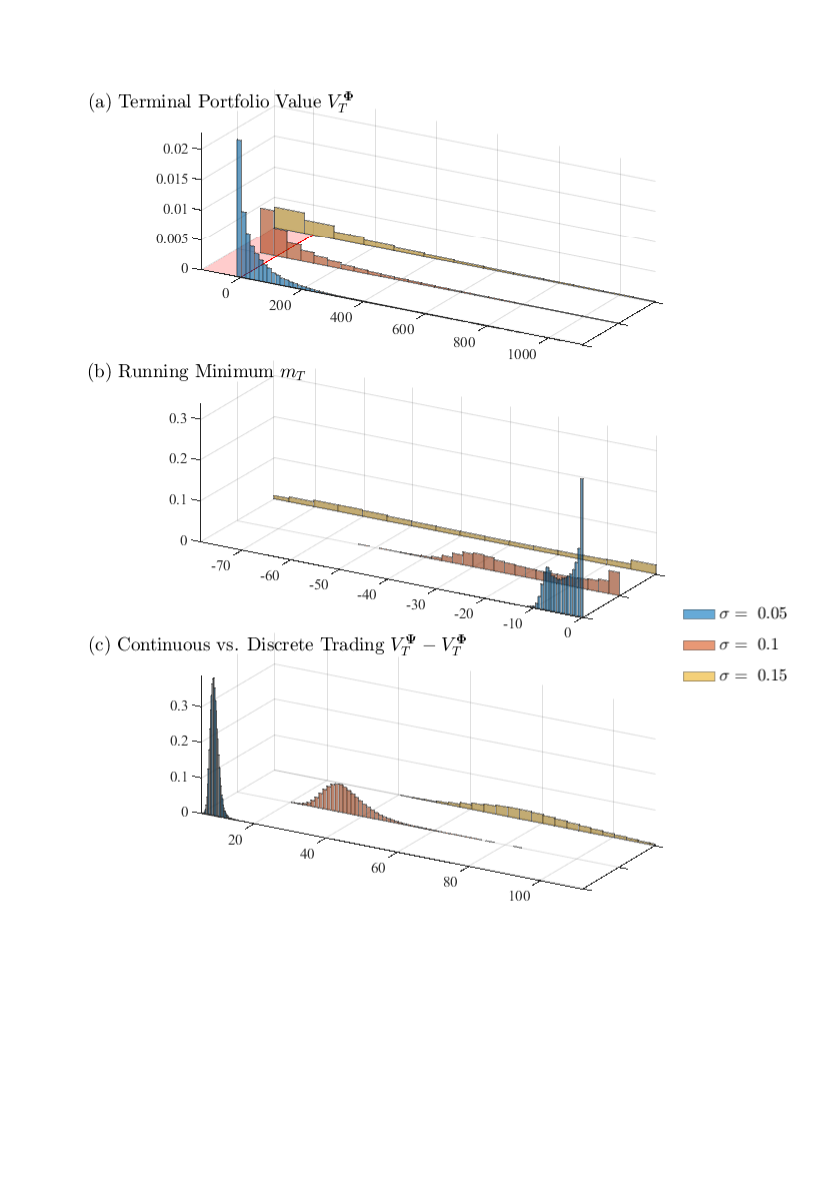}
			\end{center}		
			\scriptsize
			Similar to the Shiryaev strategy Figure \ref{Fig:Shiryaev2}, but for varying volatility values $\sigma$, this figure presents the simulated distributions of (a) the terminal portfolio value $V_T^{\stratd}$, (b) the running value process minimum $m_T$ and (c) the difference $V_T^\stratc- V_T^\stratd$ between the terminal portfolio values of continuous-time and discrete-time trading. For better visibility, the x-axis in Panels (a) and (c) (Panel (b)) ends at the 95\% quantile (starts at the 5\% quantile) of $\sigma=0.15$.
			\caption{Shiryaev portfolio value distributions for different volatilities}\label{Fig:Shiryaev4}
		\end{figure}
		
		\begin{table}[h!]
			\footnotesize
			\begin{center}
				\begin{tabular}{r|rrrrrr}
					\toprule
					$\sigma$& Mean & Median & Stand. dev.  & Min & Max & Loss prob.  \\\midrule
					$0.05$ & 46.6 \  \ {\neu (46.7)} & 19.9 & 71.4 & $-$11.3 & 1,036.6 & 0.27 \\
					$0.10$ & 109.4 {\neu (109.0)} & 25.7 & 222.8 & $-$53.5 & 4,392.8 & 0.39 \\
					$0.15$ & 218.5 {\neu (217.0)} & 41.1 & 494.9 & $-$125.6 & 11,927.1 & 0.41 \\
					\bottomrule
				\end{tabular}
			\end{center}
			\scriptsize
			Similar to Table \ref{Tab:Shiryaev_statistics}, this table reports some descriptive statistics for the simulated terminal portfolio value distributions in Panel (a) of Figure \ref{Fig:Shiryaev4}. {\neu The mean values in parentheses are approximations based on Proposition \ref{prop_conv_shiryaev}.}	
			\caption{\label{Tab:Shiryaev_sigma} Shiryaev portfolio value statistics for different volatilities}
		\end{table}
	}
	
	\afterpage{
		\begin{figure}[!ht]
			\begin{center}
				\includegraphics[trim= 0mm 46mm 0mm 10mm, clip,width=12cm]{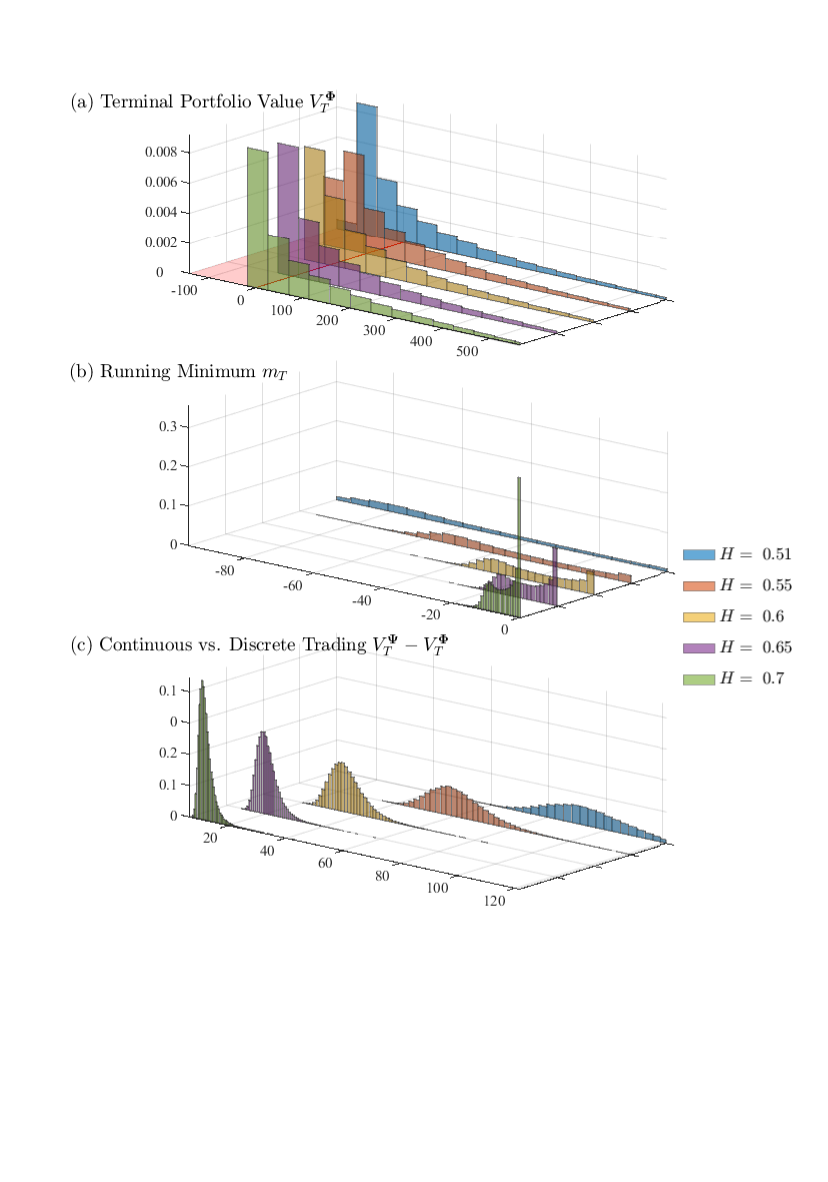}
			\end{center}
			\scriptsize
			Similar to the Shiryaev strategy Figure \ref{Fig:Shiryaev2}, but for varying Hurst coefficients $H$, this figure presents the simulated distributions of (a) the terminal portfolio value $V_T^{\stratd}$, (b) the running value process minimum $m_T$ and (c) the difference $V_T^\stratc- V_T^\stratd$ between the terminal portfolio values of continuous-time and discrete-time trading. For better visibility, the x-axis in Panels (a) and (c) (Panel (b)) ends at the 95\% quantile of $H=0.7$ and $H=0.51$, respectively (starts at the 5\% quantile of $H=0.51$).
			\caption{Shiryaev portfolio value distributions for different Hurst coefficients}\label{Fig:Shiryaev5}
		\end{figure}
		
		\begin{table}[h!]
			\footnotesize
			\begin{center}
				\begin{tabular}{r|rrrrrr}
					\toprule
					$H$& Mean & Median & Stand. dev.  & Min & Max & Loss prob. \\\midrule
					$0.51$ & 49.6 \ \ {\neu (49.0)}& $-31.7$ & 218.9 & $-141.5$ & 3,787.8 & 0.60 \\
					$0.55$ & 83.5 \ \ {\neu (83.0)}& 0.7 & 221.1 & $-91.7$ & 4,084.4 & 0.50 \\
					$0.60$ & 109.4 {\neu (109.0)}& 25.7 & 222.8 & $-53.5$ & 4,392.8 & 0.39 \\
					$0.65$ & 124.2 {\neu (123.9)}& 40.0 & 223.7 & $-31.4$ & 4,640.3 & 0.30 \\
					$0.70$ & 132.7 {\neu (132.5)}& 48.3 & 224.1 & $-18.4$ & 4,831.5 & 0.23 \\
					\bottomrule
				\end{tabular}
			\end{center}
			\scriptsize
			Similar to Table \ref{Tab:Shiryaev_statistics}, this table reports some descriptive statistics for the simulated terminal portfolio value distributions in Panel (a) of Figure \ref{Fig:Shiryaev5}. {\neu The mean values in parentheses are approximations based on Proposition \ref{prop_conv_shiryaev}.}
			\caption{\label{Tab:Shiryaev_H} Shiryaev portfolio value statistics for different Hurst coefficients}
		\end{table}
	}
	
	\paragraph{Hurst parameter \boldmath$H$} 
	In Figure~\ref{Fig:Shiryaev5} and Table \ref{Tab:Shiryaev_H}, we investigate the Hurst coefficients $H \in \{0.51,0.55,0.60,0,65,0.70\}$. Recall that $H=0.5$ implies no memory and elevating $H$ within the interval $(0.5,1)$ establishes long memory. It generates positive serial correlation with levels linked to $H$ and high even for distant lags. Shiryaev-type investors take long (short) positions in the risky asset when its price deviates from an initial state in positive (negative) direction. Thus, they can benefit directly from a trending behavior of the risky asset which is more likely under high than low $H\in(0.5,1)$. Specifically, with rising $H$, the distributions of $V_T^\stratd$ and $m_T$ relocate and reform such that the likelihood of large gains (losses) increases (decreases). Switching from $H=0.6$ to $H=0.7$, for example, raises the mean terminal value from $109.4$ to $132.7$ and lowers the loss probability from $39\%$ to $23\%$. Interestingly, this is accompanied by a sharp drop in rebalancing costs $V_T^{ \stratc}-V_T^{ \stratd}$. Hence, investors should trade assets with high $H$, which have been identified in many asset classes \cite[see][]{Hiemstra1997,Auer2016,Coakley2016}, because they make the strategy more secure and less cash-intensive with respect to the transaction account.
	
	\subsubsection{Impact of trading horizon and frequency}
	\label{subsec:horizon_frequency_Shiryaev}
	Besides a suitable risky asset, Shiryaev-type investors have to decide on the trading horizon and frequency. Thus, it is instructive to know how they affect portfolio performance.
	
\begin{figure}[ht!]
		\begin{center}
			\includegraphics[trim= 0mm 5mm 0mm 4mm, clip, width=13cm]{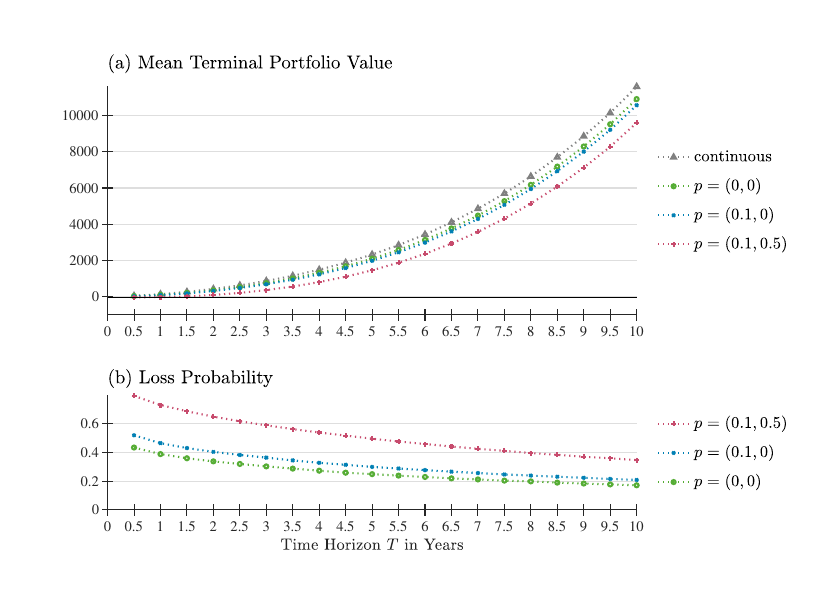}
		\end{center}
		\scriptsize	
		For 100,000 simulated scenarios of the discretized Shiryaev strategy, the basis setting of Table \ref{Tab:paramters} and a range of transaction cost values $p$, this figure plots (a) the mean of the terminal portfolio value and (b) the simulated loss probability against the trading horizon $T$. The continuous case is included as a reference.
		\caption{Shiryaev sensitivity to trading horizon}\label{Fig:Shiryaev6}
	\end{figure}

	\paragraph{Trading horizon} In a first experiment, we fix the trading frequency to daily and vary the trading horizon $T$ between $6$ months and $10$ years. As far as the remaining parameters are concerned, we use the basis setting of Table \ref{Tab:paramters} and the additional transaction cost settings of Figure \ref{Fig:Shiryaev2} and Table \ref{Tab:Shiryaev_statistics}. For each setting and trading horizon, we simulate 100,000 scenarios and report the mean terminal portfolio value and the loss probability in Figure \ref{Fig:Shiryaev6}. Panel (a) illustrates that the mean increases with the trading horizon and, except for the shortest horizons and the highest transaction costs, is positive-valued. 
	{\neu Relation \eqref{mean_theo_Shiryaev} tells us that, for large $T$, the expected terminal wealth $\E[V^\stratc_T]$ of the continuous-time Shiryaev strategy grows just like $s_0^1 \exp\{2\mu T+2\sigma^2T^{2H}\}$.} This exceeds exponential growth in $T$ because $H>0.5$. A similar behavior can be observed for the discretized strategy. Panel (b) shows that the loss probabilities initially notably decrease with $T$ and then tend to stabilize. The marginal reduction becomes less pronounced and, under proportional costs, we even reach probabilities close to 20\%. The differences between transaction cost variants also shrink with $T$ and stabilize. This can be explained by successively rising daily trading volumes eliminating minimum fee dominance. 
	
\begin{figure}[h!]
		\scriptsize
		\begin{center}
			\includegraphics[trim= 0mm 0mm 0mm 5mm, clip, width=13cm]{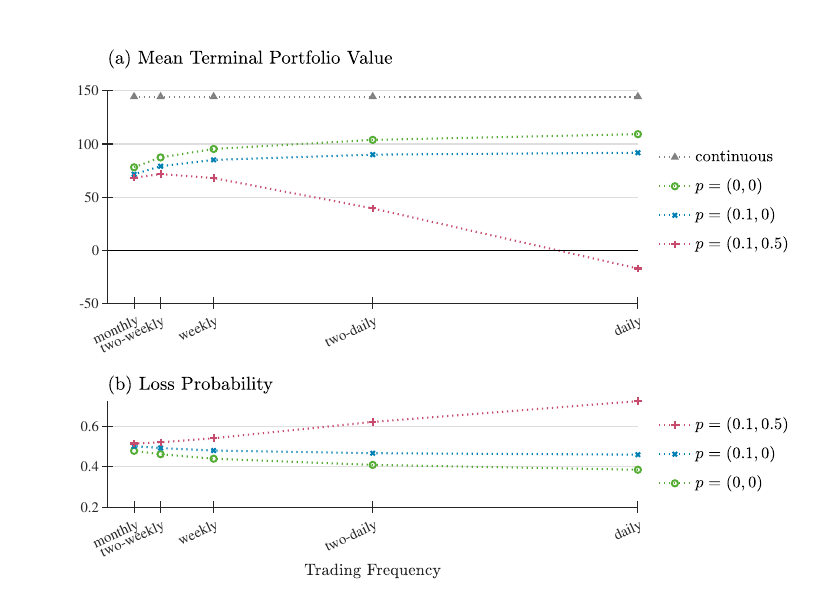}
		\end{center}
		For 100,000 simulated scenarios of the discretized Shiryaev strategy, the basis setting of Table \ref{Tab:paramters} and a range of transaction cost values $p$, this figure plots (a) the mean of the terminal portfolio value and (b) the simulated loss probability against the trading frequency. The continuous case is included as a reference.
		\caption{Shiryaev sensitivity to trading frequency}\label{Fig:Shiryaev7}
	\end{figure}

	\paragraph{Trading frequency} In a reverse second experiment, we fix the trading horizon to $T=1$ year and vary the trading frequency. We evaluate monthly, two-weekly, weekly, two-daily and daily rebalancing corresponding to $N\in\{12, 25, 50, 125, 250 \}$ trading periods per year.\footnote{We do not consider frequencies higher than daily because, in this context, our assumption of independent asset prices would no longer be realistic \cite[see][]{Malceniece2019}.} Figure \ref{Fig:Shiryaev7} presents the simulation outcomes for our three transaction cost settings, i.e., the mean terminal value and the loss probability as functions of the trading frequency. The results for trading without transaction costs and only proportional costs are similar. The mean terminal value increases with the trading frequency but there is still a clear difference to continuous-time trading.\footnote{For $p=(0,0)$, the difference disappears when the trading frequency tends to infinity.} The loss probability decreases with the trading frequency.\footnote{For $p=(0,0)$, the limiting probability is zero.} For investors facing a supplementary minimum fee, the mean terminal value sharply drops with the trading frequency and reaches a negative value for daily trading. At the same time, the loss probability rises to more than $70\%$. This feature is again caused by the scale-related relative size of proportional costs and the minimum fee.
	
	{\neu Besides supplying investment-relevant insights, simulations with rising $N$ allow us to verify the approximation quality and the convergence speed of the rebalancing cost expression derived in Proposition \ref{prop_conv_shiryaev}. For several trading frequencies, Table \ref{Tab:Shiryaev_freq} compares the simulated $\E[V_T^\stratd]$ with the approximated $\E[V_T^\stratd]$. We observe that they are very close to each other. Interestingly, this even holds for the lowest considered trading frequency, i.e., monthly rebalancing with only $N=12$ trades, where the deviation amounts to just about 3\%. Furthermore, noting that $\E[\rebalcosts(\Delta t)] = \E[V_T^\stratc-V_T^\stratd]$ and dividing the simulated $\E[V_T^\stratc-V_T^\stratd]$ by $\Delta t^{2H-1}$, we see that, in our basis setting, this ratio approaches $C=106.2$. This numerically supports our theoretical convergence results.} 
	
	\begin{table}
		\footnotesize
		{\neu
			\begin{center}
				\begin{tabular}{lr|rr|r}
					\toprule
					&& \multicolumn{2}{c|}{$\E[V_T^\stratd]$} &$(\E[V_T^\stratc-V_T^\stratd])/\Delta t^{2H-1}$   \\
					Frequency & $N$& simulated  & approximated & \multicolumn{1}{c}{simulated}    \\
					\midrule
					Monthly      & 12  &   77.1 &  79.6 & 109.6 \hspace*{3em}   \\       
					Two-weekly   & 25  &   87.4 &  88.4 & 109.2 \hspace*{3em}  \\
					Weekly       & 50  &   95.3 &  95.6 & 108.0  \hspace*{3em} \\
					Two-daily    & 125 &  103.9 & 103.4 & 107.2 \hspace*{3em}  \\
					Daily        & 250  & 109.4 & 109.0 & 106.6 \hspace*{3em}  \\
					\midrule
					& $\infty$ & 144.7  & { $\E[V_T^\stratc] =$} 144.2  & $C = 106.2$ \hspace*{3em} \\   
					\bottomrule
				\end{tabular}
			\end{center}
		}
		\scriptsize
		{\neu For several trading frequencies of the Shiryaev strategy and the basis setting of Table \ref{Tab:paramters}, this table contrasts the simulated and the approximated values of $\E[V_T^\stratd]$ where the latter are obtained via Proposition \ref{prop_conv_shiryaev}. The numbers in the last line represent the simulated and exact values of $\E[V_T^\stratc]$ taken from Table \ref{Tab:Shiryaev_statistics}. The last table column divides the simulated value of $\E[V_T^\stratc-V_T^\stratd]$ by $\Delta t^{2H-1}$. $C$ is the rate of convergence in Proposition \ref{prop_conv_shiryaev}.}  
		\caption{\label{Tab:Shiryaev_freq} {\neu Shiryaev mean approximation}}
	\end{table}    
	
	\subsection{Discretized Salopek strategy}\label{sec:Analysis_Salopek} 
	
	\subsubsection{\neu General strategy behavior}
	\label{subsec:Salopek_general}
\begin{figure}[h!]
		\begin{center}
			\includegraphics[trim= 10mm 10mm 0mm 10mm, clip, width=12cm]{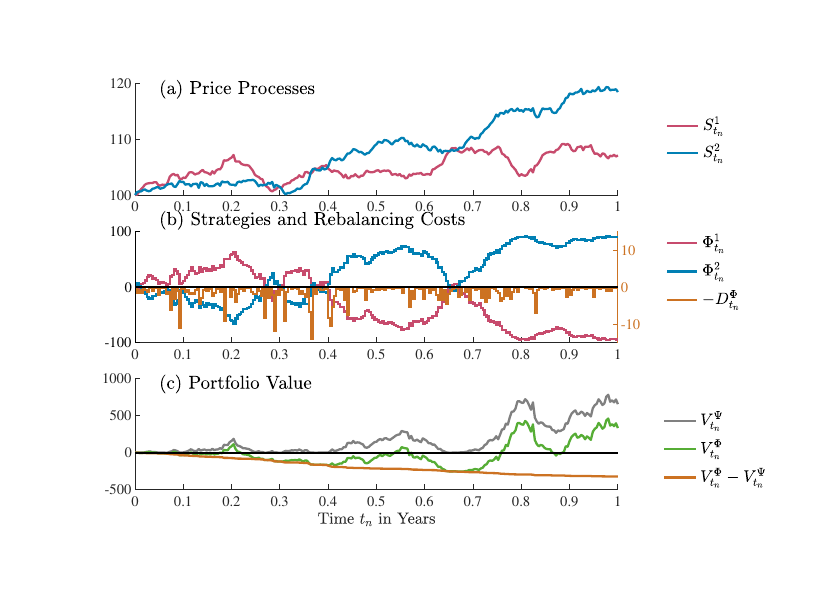}
		\end{center}
		\scriptsize
		For the discretized Salopek strategy and our basis setting of Table \ref{Tab:paramters}, this figure plots a typical simulation result. Panel (a) shows the realized prices $S^1_{t_n}$ and $S^2_{t_n}$ of the two risky assets. Panel (b) illustrates the strategy holdings $\Phi^1_{t_n}$ and $\Phi^2_{t_n}$ for these assets. Furthermore, it contains the negative rebalancing costs $-D_{t_n}^{\Phi}$. Finally, Panel (c) reports the portfolio value $V_{t_n}^{\stratd}$ of the strategy. It is supplemented by the value process $V_{t_n}^{\stratc}$ of continuous-time trading and the difference between both portfolio values which, except for terminal time $t_N=T= 1$, is equal to the holdings $\Phi^3_{t_n}$ in the transaction account. 	
		\caption{Exemplary realization of the discretized Salopek strategy}\label{Fig:Salopek1}	
	\end{figure}

	We now turn to the Salopek strategy trading only risky assets and put special emphasis on a simple and practically appealing specification with $d=2$  assets.\footnote{For the strategy to work, the prices of the two assets should not be perfectly correlated. This is ensured by our independence assumption of Section \ref{subsec:ContinuousModel}.} Following our approach for the Shiryaev strategy, Figure \ref{Fig:Salopek1} starts by presenting a typical simulated realization in the basis setting of Table \ref{subsec:ModelParameter}. This means that we plot the prices $S^1_{t_n}$ and $S^2_{t_n}$, the asset holdings $\Phi^1_{t_n}$ and $\Phi^2_{t_n}$, the negative rebalancing costs $-D_{t_n}^{\Phi}$ as well as the discrete and continuous strategy value processes $V_{t_n}^{\stratd}$ and $V_{t_n}^{\stratc}$ including their differences $V_{t_n}^{\stratd}-V_{t_n}^{\stratc}$. 
	
	In line with Proposition \ref{prop_Salopek}, we see that the investor is always long (short) in the asset with the higher (lower) price. Because the continuous-time value \eqref{eq:TermValueSalopek} tells us that $V_t^{\stratc}= M_\beta(S_t)- M_\alpha(S_t)\ge 0$, the properties of the $a$-order power mean $M_a(.)$ imply that the portfolio value of the Salopek strategy is all the greater the more the prices of the two assets deviate from each other. If they coincide, we have $M_\beta(S_t)= M_\alpha(S_t)$ and consequently a $V_t^{\stratc}$ of zero. These features are comparable to the Shiryaev strategy.
	
	At first glance, it appears that there are also strong rebalancing cost similarities between the Shiryaev and Salopek strategies. For the Shiryaev strategy, we have shown that the rebalancing costs $D_{t_n}^{\Phi}$ are strictly positive for $n=1,\ldots,N-1$ (see Proposition \ref{Shiryaev_rebalancing}). Figure \ref{Fig:Salopek1} suggests the same property for the Salopek strategy.\footnote{{\neu Both strategies have in common that their rebalancing costs tend to be highest when asset prices experience sharp changes and price paths cross.}} However, this does not hold in general. It can be verified via experiments with different choices of $(\alpha,\beta)$ that $D_{t_n}^{\Phi}$ may be negative for some $n$ (see, for example, Figure \ref{Fig:SalopekPositiveDifferences} of the appendix). Thus, in contrast to the Shiryaev strategy, the portfolio value $V_{t_n}^{\stratd}$ of the discretized Salopek strategy can exceed the portfolio value $V_{t_n}^{\stratc}$ of its continuous-time counterpart.
	
	\subsubsection{Impact of {\neu time discretization and} transaction costs}\label{subsec:transaction_Salopek}
\begin{figure}[ht!]
		\begin{center}
			\includegraphics[trim= 0mm 46mm 0mm 10mm, clip, width=12cm]{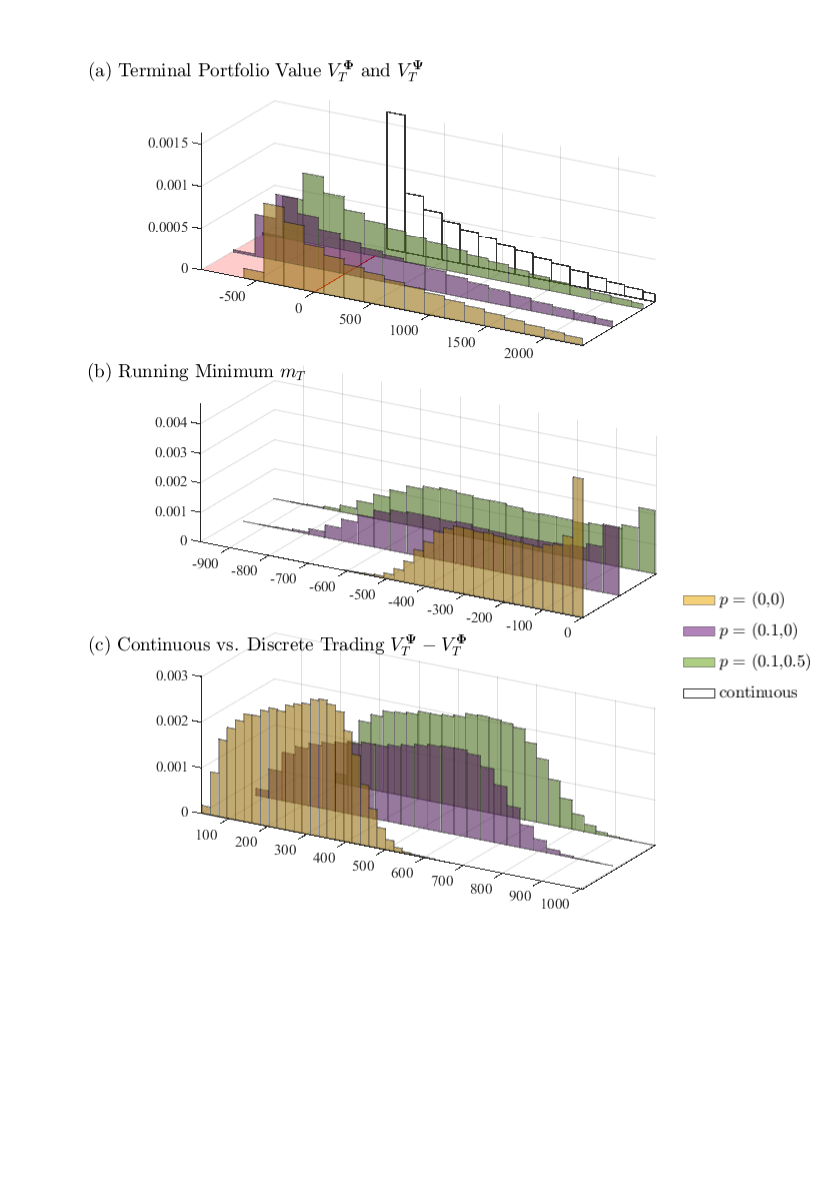}
		\end{center}
		\scriptsize
		For 100,000 simulated scenarios of the discretized Salopek strategy, the basis setting of Table \ref{Tab:paramters} and a range of transaction cost values, this figure presents various portfolio value distributions. Panel (a) shows the distributions of the terminal value $V_T^{\stratd}$ of discrete-time trading with transaction costs $p = (p_1,p_2)$ where $p_1$ reflects proportional costs (in percent) and $p_2$ is a minimum fee (in monetary units). The loss region with negative terminal values is highlighted by a red floor. The distribution of the terminal value $V_T^{\stratc}$ of continuous-time trading is also included. Panel (b) contains the distributions of the running minimum $m_T$ of the discrete value processes, i.e., the worst-case portfolio values in the investment horizon. Finally, the distributions in Panel (c) refer to the terminal difference $V_T^\stratc- V_T^\stratd$ between continuous-time and discrete-time trading. For better visibility, the x-axis in Panel (a) is cut off after the 95\% quantile of $p=(0,0)$.
		\caption{Salopek portfolio value distributions for different transaction costs}
		\label{Fig:Salopek2}
	\end{figure}

	Uniting all 100,000 simulation scenarios and charging transaction costs in the Salopek strategy, Figure \ref{Fig:Salopek2} presents the distributions of the terminal portfolio values $V_T^{\stratc}$ and $V_T^{\stratd}$, the running minimum $m_T$ and the difference $V_T^\stratc-V_T^\stratd$. In addition, Table \ref{Tab:Salopek_statistics} reports summary statistics for the terminal value distributions.
	
	Similar to the Shiryaev case, discretization and transaction costs expand the negative distribution support of $V_T^{\stratd}$ and $m_T$ and increase $V_T^\stratc-V_T^\stratd$. However, the Salopek strategy differs in notable aspects. First, without transaction costs, the means of $V_T^{\stratc}$ and $V_T^{\stratd}$ are $806.3$ and $534.1$, respectively. Because $534.1$ covers just about 66\% of $806.3$, the Salopek strategy exhibits larger discretization shrinkage than the Shiryaev strategy. {\neu The theoretical counterpart of the latter value, delivered by the numerical integration of Remark \ref{rem_distribution_Salopek}, is $805.9$. The former quantity can be approximated via Proposition \ref{prop_conv_salopek} which derives a rebalancing cost expansion (under zero transaction costs) with convergence features similar to the Shiryaev Proposition \ref{prop_conv_shiryaev}. It quite accurately yields 537.7 (see also Table \ref{Tab:Salopek_freq}).} Second, turning to $p=(0.1,0)$, the terminal mean and loss probability are $349.7$ and $47\%$, respectively. These values are higher than for the Shiryaev strategy but must be put into the perspective that the Salopek strategy drains the transaction account more significantly than the Shiryaev strategy. Finally, for $p=(0.1,0.5)$, we observe a terminal mean of $303.0$ accompanied by a loss probability of $48\%$. While, in the Shiryaev case, the minimum fee causes a negative mean and a very high loss probability, this does not occur for the Salopek strategy because of its larger trading volumes. Overall, transaction costs do not diminish the performance of the discretized strategy to a point of financial futility. 
	
	{\neu
		\begin{prop}[Expected cumulated Salopek rebalancing costs]\label{prop_conv_salopek}	Assume a market with an equidistant time discretization of step size $\Delta t>0$ and consider the discrete-time Salopek strategy with parameters $\alpha,\beta \in \R\setminus \{0\}, \alpha<\beta$. Then, for $\Delta t\to 0$, the expectation of its cumulated rebalancing costs $\rebalcosts=\rebalcosts(\Delta t)=\sum_{n=1}^{N} D_{t_n}^{\stratd}$ given in \eqref{def_cum_rebalancing} satisfies
			\begin{align}
				\E\big[\rebalcosts(\Delta t) \big]&  = C\, \Delta t^{2H_{\min}-1} + o(\Delta t^{2H_{\min}-1}) \quad\text{with }\quad C= { C(\alpha,\beta) = \overline{C}(\beta)-\overline{C}(\alpha)},\\
				\text{where }\quad  { \overline{C}(a)} &=  \frac{a-1}{2d} \int_0^T 
				\E\bigg[M_a(S_{t})\sum\limits_{i\in \Hminindex} (\sigma^i)^2 \Big(\frac{S_{t}^i}{M_a(S_{t})}\Big)^a \bigg(1-\frac{1}{d}\Big(\frac{S_{t}^i}{M_a(S_{t})}\Big)^a\bigg)\bigg] \,dt,\\ 
				H_{\min} &=\min\{H^1,\ldots,H^d\},~\Hminindex=\{i\in\{1,\ldots,d\}: H^i=H_{\min}\}.
			\end{align}	
		\end{prop}
		\vspace{-0.2cm}
		\begin{proof}
			See Appendix \ref{prop_conv_salopek_proof}.   
		\end{proof}
	}
	
	\begin{table}[ht!]
		\footnotesize
		\begin{center}
			\begin{tabular}[t]{lr|dd|ddddd|e}
				\toprule
				Strat.& Transact.& \mc{Mean} & \mcl{ Stand.} &  & \multicolumn{3}{c}{ Quantiles}  &  & \mc{Loss}  \\   	
				& costs $p$& & \mcl{dev.}    & \mc{Min} & \mc{5\%} & \mc{Median}   & \mc{95\%}& \mcl{Max} & \mc{prob.} \\
				\midrule
				$\stratc$ & none & 806.3 & 817.7 & 0.0 & 6.1 & 551.3 & 2,462.3 & 6,813.7 & 0.00 \\
				&  &  (805.9) & (813.5) &  (0.0) &  (6.7) &  (556.4) &  (2,458.6) &  (\infty) &  (0.00) \\
				\midrule
				$\stratd$ & $(0,0)$  &534.1 & 893.6 & -609.3 & -393.6 & 273.9 & 2,322.1 & 6,682.6 & 0.37 \\
				& $(0.1,0)$ &349.7 & 940.1 & -962.7 & -660.3 & 85.2 & 2,219.2 & 6,591.0 & 0.47 \\
				& $(0.1,0.5)$ &303.0 & 920.3 & -977.7 & -680.7 & 42.4 & 2,136.0 & 6,492.0 & 0.48 \\
				\bottomrule
			\end{tabular}
		\end{center}
		\scriptsize
		This table reports some descriptive statistics for the simulated terminal portfolio value distributions in Panel (a) of Figure \ref{Fig:Salopek2}. Besides the mean and standard deviation, we compute the minima and maxima as well as selected quantiles. Furthermore, we present the simulated loss probability, i.e., the proportion of negative terminal portfolio values. {\neu The numbers in parentheses for the continuous-time portfolio represent theoretical values obtained via the numerical integration of Remark \ref{rem_distribution_Salopek}.}
		\caption{Salopek portfolio value statistics for different transaction costs}
		\label{Tab:Salopek_statistics} 
	\end{table}
	
	{\neu     
		\begin{bem}\label{rem_conv_salopek} For the special case of $d$ assets with identical Hurst coefficients and volatilities, i.e., $H^i=H, \sigma^i=\sigma, i=1,\ldots,d$, which we consider in our basis setting, the expression defining $\overline{C}(a)$ in Proposition \ref{prop_conv_salopek} can be simplified to
			\begin{align}
				\overline{C}(a) &=  \frac{(a-1)\sigma^2}{2} \int_0^T 
				\E\Big[M_a(S_{t})  \Big(1-\frac{1}{d}\Big(\frac{M_{2a}(S_{t})}{M_a(S_{t})}\Big)^{2a}\Big)\Big] \,dt.
			\end{align}
			Here, the equality of the drifts $\mu^1,\ldots,\mu^d$ is not required.
		\end{bem}	
	} 
	
	\subsubsection{Impact of Hurst parameters}
\begin{figure}[h!]
		\begin{center}
			\includegraphics[trim= 0mm 10mm 0mm 5mm, clip, width=12cm]{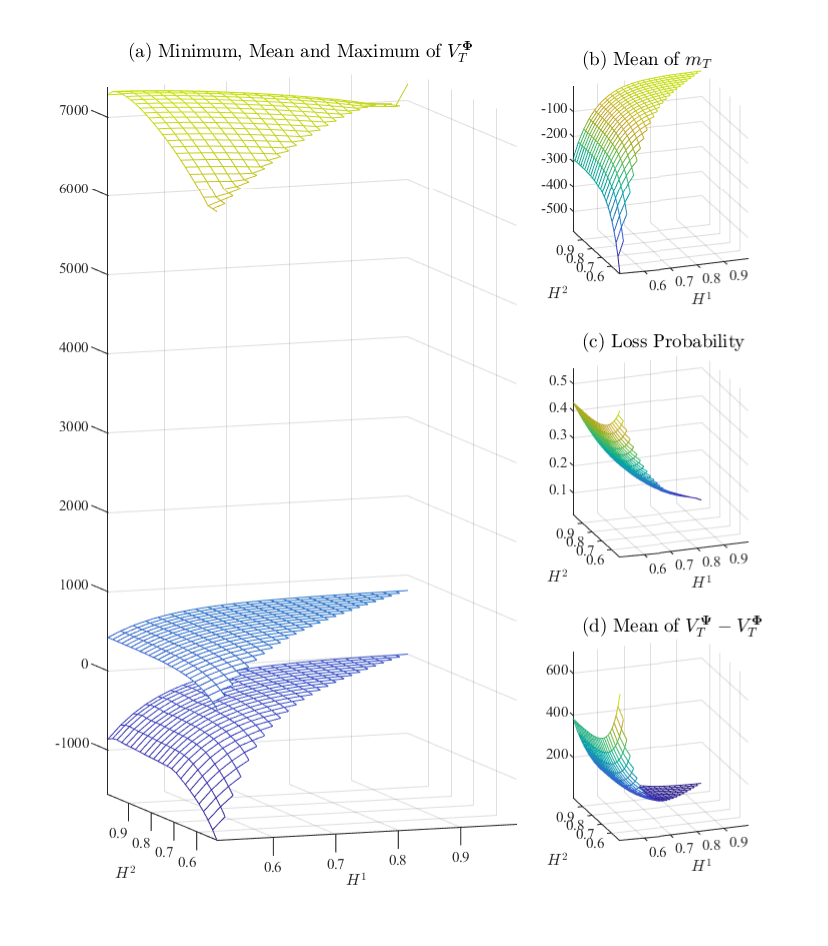}
		\end{center}		
		\scriptsize
		For 100,000 simulated scenarios of the discretized Salopek strategy, the basis setting of Table \ref{Tab:paramters} and a range of Hurst parameter values $H^1,H^2\in[0.51,0.99]$ with $H^1\le H^2$, this figure characterizes the corresponding distributions of the terminal portfolio value $V_T^{\stratd}$. Panel (a) shows the minimum, mean and maximum of $V_T^{\stratd}$. Panel (b) plots the mean of the running minimum $m_T$. Panels (c) and (d) cover the simulated loss probability and the mean of the difference $V_T^{\stratc}-V_T^{\stratd}$ between continuous and discrete trading, respectively. 
		\caption{Salopek sensitivity to Hurst coefficients}\label{Fig:Salopek3}
	\end{figure}

	Our basis setting assumes that the Hurst coefficients of the traded assets are identical, i.e., $H^1=H^2$. Because this is not a necessary requirement for strategy execution, we also investigate Hurst values of different magnitudes. Given the complexity of this exercise, Figure~\ref{Fig:Salopek3} presents its results, i.e., the characteristics of the terminal portfolio value distributions, in three-dimensional form. For $H^1,H^2\in[0.51,0.99]$ with $H^1\le H^2$, Panel (a) plots the minimum, mean and maximum of $V_T^{\stratd}$.\footnote{The outcomes for $H^1>H^2$ follow by symmetry.} Panels (b), (c) and (d) cover the mean of the running minimum $m_T$, the loss probability and the mean of $V_T^{\stratc}-V_T^{\stratd}$, respectively.
	
	Higher persistence pushes terminal values, limits the risk of loss and reduces capital infusions. If both $H^1$ and $H^2$ approach their limit value one, the loss probability and the mean of the running minimum are drawn to zero. Put differently, even though discretization causes $V_T^\stratc-V_T^\stratd \neq 0$, the discretized Salopek strategy converges to an almost perfect arbitrage strategy. This can be explained by the limiting behavior of the fBm $B^{H}$  for $H\to 1$. 
	In this situation, \eqref{fBm_covariance} implies for the covariance $\textrm{Cov}(B_t^{H},B_s^{H}) \to ts$ such that $B_t^{H}$ and $B_s^{H}$ are perfectly positively correlated for all $t,s> 0$ and it can be deduced that $B_t^{H}=tB_1^{H}$. Because the fBm is a centered Gaussian process, we obtain $B^{H^i}_t=\gauss^i t$ with  independent standard Gaussian random variables $\gauss^i, i=1,2$. In this case, (\ref{eq:BlackScholes}) delivers asset prices $S_t^i=s_0^i \exp\{\mu^it+\sigma^i\gauss^it\}$. Because their paths are exponential functions with growth rate $\mu^i+\sigma^i\gauss^i$, the quantities $\gauss^i$ are the only source of uncertainty. However, they are unveiled to the investor with the asset price observations at the first trading time $t_1$. This means that, after $t_1$, future asset prices are completely known. If the growth rate of $S^1$ is above (below) the one of $S^2$, we have $S^1_t>S^2_t$ ($S^1_t<S^2_t$) for all $t\in[0,T]$. Thus, a long position in the asset with the larger price and a short position in the other is an obviously risk-free strategy.
	
	\subsubsection{Impact of strategy parameters}
	\label{subsec:Salopek_parameters}
	While Shiryaev-type investors have access to a unique trading rule, Salopek investors are confronted with a family of rules parameterized by $(\alpha,\beta)$. Consequently, they need to choose a suitable tuple $(\alpha,\beta)$ in practical applications. In the continuous-time case, \eqref{eq:TermValueSalopek} and \eqref{eq:mean_monoton} imply that the maximum portfolio value $V_{t}^{\stratc,\max}=\max\left(S_{t}^1,S_{t}^2\right)-\min\left(S_{t}^1,S_{t}^2\right)$ is attained for the limiting tuple $(\alpha,\beta)=(-\infty,\infty)$. With this setup, the strategy representation \eqref{eq:continuousSalopekStrategy} tells us that $\stratc$ is a \textit{buy-and-hold} strategy with a long position in the high-priced asset financed by short selling the low-priced asset as long as the sign of $S^1-S^2$ is unchanged. If a change occurs, i.e., if the price paths cross, the asset roles simply reverse.\footnote{A graphical illustration of this strategy can be found in Figure~\ref{Fig:SalopekStrategyLimit} of the appendix.} 
	
	\begin{figure}[ht!]
		\begin{center}
			\includegraphics[trim= 0mm 12.5mm 0mm 5mm, clip, width=11.5cm]{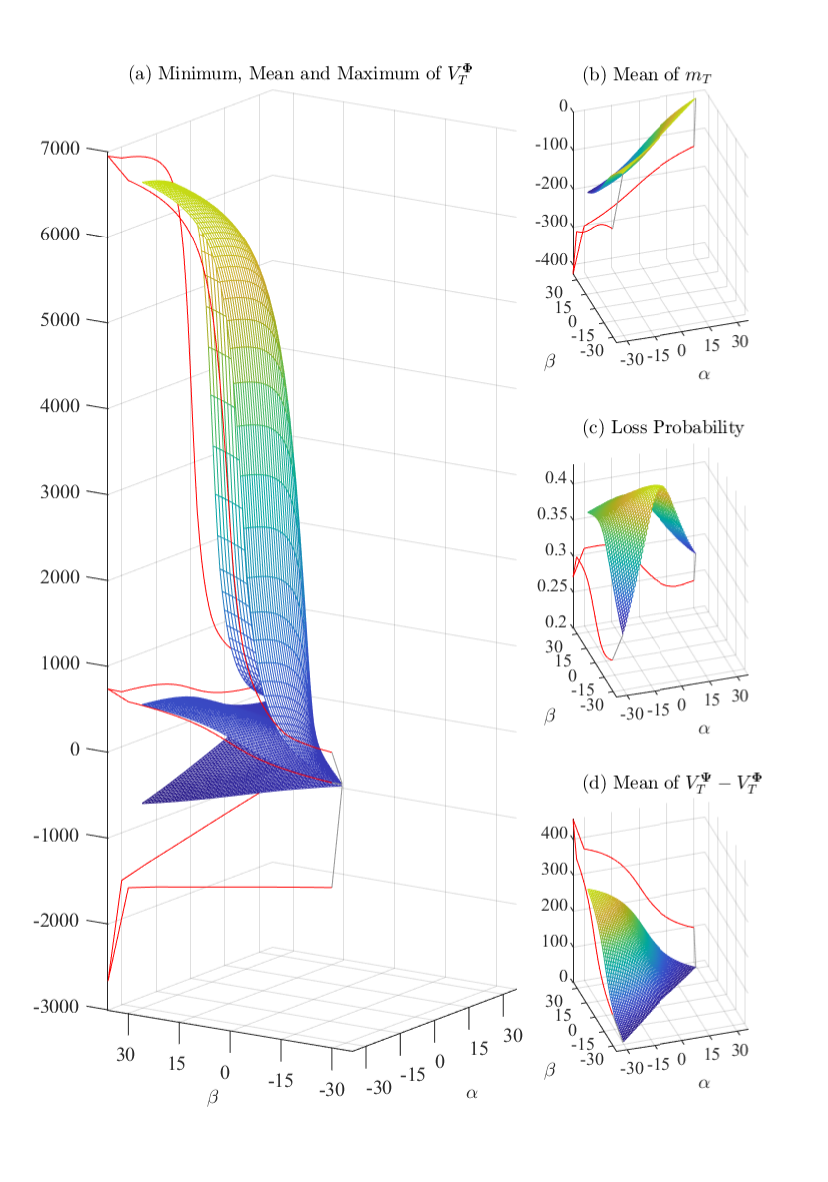}
		\end{center}
		\scriptsize
		For 100,000 simulated scenarios of the discretized Salopek strategy, the basis setting of Table \ref{Tab:paramters} and a range of strategy parameter values $\alpha\in[-30,29]$ and $\beta\in [\alpha+1,30]$, this figure characterizes the corresponding distributions of the terminal portfolio value $V_T^{\stratd}$. Panel (a) shows the minimum, mean and maximum of $V_T^{\stratd}$. Panel (b) plots the mean of the running minimum $m_T$. Panels (c) and (d) cover the simulated loss probability and the mean of the difference $V_T^{\stratc}- V_T^{\stratd}$ between continuous and discrete trading, respectively. The red lines represent the results for the limiting cases $\alpha=-\infty$ and $\beta=\infty$.
		\caption{Salopek sensitivity to strategy parameters}\label{Fig:Salopek4}
	\end{figure}
	
	Although the infinite setup is appealing from a theoretical point of view, the question arises whether it also maxes out in a discrete environment. Rebalancing costs may vary with $(\alpha,\beta)$ and suggest a different optimal parameter choice. To provide an answer, Figure~\ref{Fig:Salopek4} plots our set of previously used portfolio value characteristics against the parameters $\alpha\in[-30,29]$ and $\beta\in [\alpha+1,30]$. For $\alpha=\beta$, the Salopek positions and portfolio value are zero because we are not invested in any risky asset. With growing difference $\beta -\alpha$, the means of $V_T^{\stratd}$ and  $V_T^{\stratc}-V_T^{\stratd}$ increase. They reach their maxima for the limit $\beta=-\alpha=\infty$. Thus, even though the rebalancing costs are also at their maximum, continuous-time and discrete-time trading both max in the same limiting case. With respect to the loss probabilities, we observe values between $23\%$ and $43\%$. The highest probability arises for the tuple $(\alpha,\beta)=(0,1)$, the lowest for $(20,\infty)$. In the case of $\beta=-\alpha=\infty$, we have $27\%$.
	
	{\neu Investors basically have two well-known options when setting their individual $(\alpha,\beta)$. First, they may tune the parameters by determining the values that optimize a given target function (or a set of target functions). For example, the reward earned per unit of investment risk could be maximized \cite[see][]{Xiang2024}. Such an endeavor yields diverse results because the literature offers a wide variety of alternative reward-to-risk ratios \cite[see][]{Rachev2007,Eling2007}. Second, and very popular in practice, a sequential approach is feasible. That is, investors use visualizations like Figure \ref{Fig:Salopek4} to check whether a desired reward goes along with personally acceptable risk and budget levels \cite[see][]{Bodie2009}. If it does, the corresponding parameters are chosen. Otherwise, the reward must be adjusted to meet risk and budget preferences.}
	
	\subsubsection{Impact of trading horizon and frequency}
	\label{subsec:horizon_frequency_Salopek}
	To complete our analysis of the Salopek strategy, we study its sensitivity to different trading horizons and frequencies and compare the findings to the Shiryaev strategy. 
	
\begin{figure}[ht!]
		\scriptsize
		\begin{center} 
			\includegraphics[trim= 0mm 5mm 0mm 4mm, clip,width=12cm]{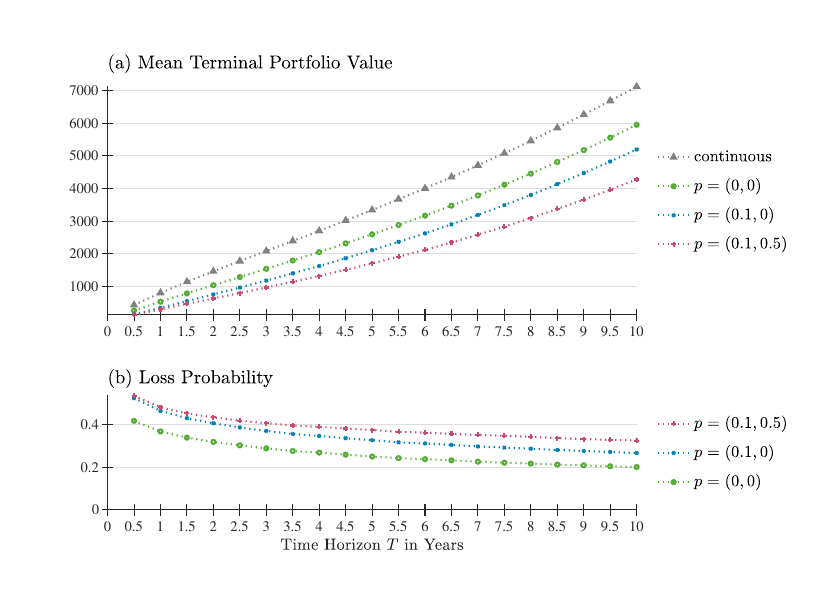}
		\end{center}
		For 100,000 simulated scenarios of the discretized Salopek strategy, the basis setting of Table \ref{Tab:paramters} and a range of transaction cost values $p$, this figure plots (a) the mean of the terminal portfolio value and (b) the simulated loss probability against the trading horizon $T$. The continuous case is included as a reference.
		\caption{Salopek sensitivity to trading horizon}\label{Fig:Salopek6}
	\end{figure}

	\paragraph{Trading horizon} Again, we start by enlarging the trading horizon $T$ and upholding a daily trading frequency. For our three transaction cost variants, Figure \ref{Fig:Salopek6} displays the mean terminal portfolio value and the loss probability as functions of $T$. For both $\stratc$ and $\stratd$, Panel (a) suggests that the growth of the mean terminal values in the first ten years is only slightly faster than linear. This is in contrast to the Shiryaev strategy where we detected faster than exponential growth. While, for smaller $T$, the means of the Salopek strategy are larger than those of the Shiryaev strategy, the latter surpass the former for higher $T$. In particular, for $T=10$, the Shiryaev values exceed the Salopek values nearly twofold. In our drifting environment, this can be linked to the fact that the Shiryaev asset tends to deviate further from its initial price than the Salopek assets deviate from each other. Panel (b) shows that the loss probabilities decrease with $T$. While the Salopek strategy has some comparative advantages in the short run and the minimum fee setting, the Shiryaev strategy stands out in the long run. In particular, for $T=10$ and proportional costs, the former is not as close to 20\% as the latter. Moreover, with rising $T$, the Salopek probabilities related to minimum fee transaction costs do not approach those of purely proportional costs.
	
\begin{figure}[h!]
		\begin{center} 
			\includegraphics[trim= 0mm 0mm 0mm 5mm, clip,width=12cm]{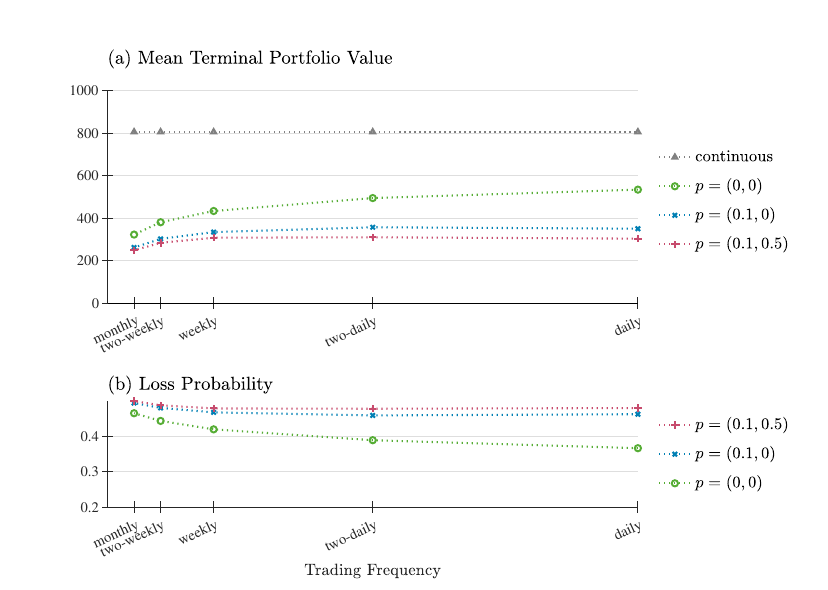}
		\end{center}	
		\scriptsize
		For 100,000 simulated scenarios of the discretized Salopek strategy, the basis setting of Table \ref{Tab:paramters} and a range of transaction cost values $p$, this figure plots (a) the mean of the terminal portfolio value and (b) the simulated loss probability against the trading frequency. The continuous case is included as a reference.
		\caption{Salopek sensitivity to trading frequency}	\label{Fig:Salopek7}
	\end{figure}

	\paragraph{Trading frequency} Figure \ref{Fig:Salopek7} sets diverse trading frequencies within a locked trading horizon of $T=1$ year. The resulting mean terminal portfolio values in Panel (a) imply that, even in the presence of transaction costs, higher trading frequencies are beneficial for investors. In contrast to the Shiryaev strategy, the impact of minimum fees at high frequencies is almost negligible. The loss probabilities in Panel (b) also recommend more frequent trading. The value of $37\%$ for fee-less daily trading illustrates once more that unavoidable rebalancing costs in discrete-time trading induce losses that prevent the strategy from reaching the zero probability limit of an infinite trading frequency.
	
	{\neu Finally, similar to the Shiryaev investigation, Table \ref{Tab:Salopek_freq} studies the approximation accuracy and the convergence speed of the expansion in Proposition \ref{prop_conv_salopek}. The proxy proves to be very useful because the deviations between simulated and approximated values range from roughly 6\% at low frequency ($N=12$) to about 0.25\% at high frequency ($N=250$). In addition, the proposed convergence rate $C$ and order $2H-1$ are nicely confirmed by our simulation exercise.}
	
	\begin{table}[h!]
		\footnotesize
		{\neu
			\begin{center}
				\begin{tabular}{lr|rr|r}
					\toprule
					&& \multicolumn{2}{c|}{$\E[V_T^\stratd]$} &$(\E[V_T^\stratc-V_T^\stratd])/\Delta t^{2H-1}$   \\
					Frequency & $N$& simulated  & approximated & \multicolumn{1}{c}{simulated}  \\
					\midrule
					Monthly    & 12  &  323.2 & 304.6 &   794.1 \hspace*{3em}  \\       
					Two-weekly & 25  &  381.3 & 373.0 &   809.1 \hspace*{3em} \\
					Weekly     & 50  &  434.3 & 429.0 &   813.5 \hspace*{3em} \\
					Two-daily  & 125 &  494.2 & 492.1 &   817.1 \hspace*{3em} \\
					Daily      & 250 &  534.1 & 532.7 &   821.3 \hspace*{3em}  \\   
					\midrule
					& $\infty$ &   806.3  & { $\E[V_T^\stratc] =$} 805.9  & $C = 824.0$ \hspace*{3em}  \\   
					\bottomrule
				\end{tabular}
			\end{center}
		}
		\scriptsize
		{\neu For several trading frequencies of the Salopek strategy and the basis setting of Table \ref{Tab:paramters}, this table contrasts the simulated and the approximated values of $\E[V_T^\stratd]$ where the latter are obtained via Proposition \ref{prop_conv_salopek}. The numbers in the last line represent the simulated and exact values of $\E[V_T^\stratc]$ taken from Table \ref{Tab:Salopek_statistics}. The last table column divides the simulated value of $\E[V_T^\stratc-V_T^\stratd]$ by $\Delta t^{2H-1}$. $C$ is the rate of convergence in Proposition \ref{prop_conv_salopek}.}
		\caption{\label{Tab:Salopek_freq} {\neu Salopek mean approximation}}
	\end{table}

	{\neu 
		\section{Empirical illustration}
		\label{sec:emp}
		To complete our study, we leave the simulation setup with artificially generated data and shed some light on the implementation and performance of the Shiryaev and Salopek strategies in empirical data. Specifically, we obtain an illustrative data set from Refinitiv Eikon which spans from January 2013 to December 2023 and covers daily price information of the 1,689 full-history stocks listed in the 41 country subindices of the Datastream Emerging Market Index.\footnote{{\neu The request mnemonic for this data is G\#LTOTMKEK0424. In additional calculations with developed market equities (i.e., the constituents of the DAX 40, EURO STOXX 50, FTSE 100, and S\&P 500), we attained notably weaker results.}} In this asset universe, we then conduct an rolling window analysis which is an established standard in investment research and practice because it avoids look-ahead bias with respect to asset selection criteria and can take into account the time-varying nature of memory effects and other return properties \cite[see][]{Mehlitz2024}.
		
		We opt for annually rolling a five-year data window. This means that, at the beginning of each year between 2018 and 2023, we have to use data of the preceeding five years to determine which stocks are eligible to enter our strategies. Specifically, we need to know which of them are compatible with a fBm exhibiting $H>0.5$.\footnote{{\neu As stated in Appendix \ref{sec:mfBm}, our specific strategies only hold for the case $H>0.5$ where the fBm features zero quadratic variation. Investors interested in exploiting $H<0.5$, where the quadratic variation of the fBm is infinite \cite[see][]{Biagini2008}, may turn to the strategies of \cite{Garcin2022} and \cite{Xiang2024}.}} Unfortunately, the current state of the statistical literature does not supply procedures that can reliably test for fBm behavior, i.e., simultaneously verify the validity of the model and estimate its parameters \cite[see][]{Sikora2018}. Therefore, we follow a heuristic approach popular in empirical research. That is, we deduce the fBm property from previous studies persuasively suggesting predictable variation in emerging market equities \cite[see][]{Wilcox2008} and consider a measured $H$ to be economically relevant if it exceeds $0.6$ \cite[see][]{Batten2013}. To estimate the Hurst coefficient of a stock with the window data, we use the traditional rescaled range method \cite[see][]{Weron2002} because it is fairly robust to short time-series, short-term memory and non-normality \cite[see][]{Chamoli2007,Kristoufek2012}.\footnote{{\neu Note that rescaled range analysis should utilize logarithmic instead of simple returns \cite[see][]{Auer2016}.}} Based on the obtained estimates, we exclude stocks with $H$ values below $0.6$. The remaining stocks are then rebased to one and traded for a one-year horizon and on a daily frequency. As far as the strategy specifications are concerned, we use $p = (0.1,0)$, $\gamma = 10^2$ and $(\alpha,\beta) = (-30,30)$. While the Shiryaev strategy is implemented for each stock, the Salopek strategy considers all possible pairs of stocks.  
		
			\begin{figure}[ht!]
				\begin{center} 
					\includegraphics[trim= 0mm 0mm 0mm 0mm, clip,width=12cm]{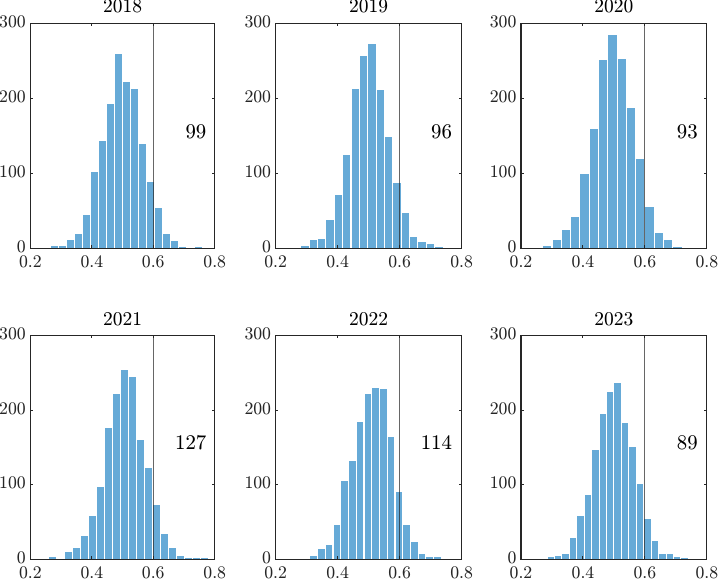}
				\end{center}	
				\scriptsize
				{\neu For each investment period in our sample, this figure displays the distribution of emerging market equity Hurst coefficient estimates. Estimation is performed by traditional rescaled range analysis and a daily return data window covering the five years prior to the investment period. The numbers within the figures show how many stocks satisfy $H\geq0.6$, i.e., enter our investment strategies.}
				\caption{{\neu Empirical Hurst coefficient values}}\label{Fig:EmpHurst}
			\end{figure}
			
			\begin{figure}[h!]
				\vspace{0.25cm}
				\begin{center} 
					\includegraphics[trim= 0mm 0mm 0mm 0mm, clip,width=14.25cm]{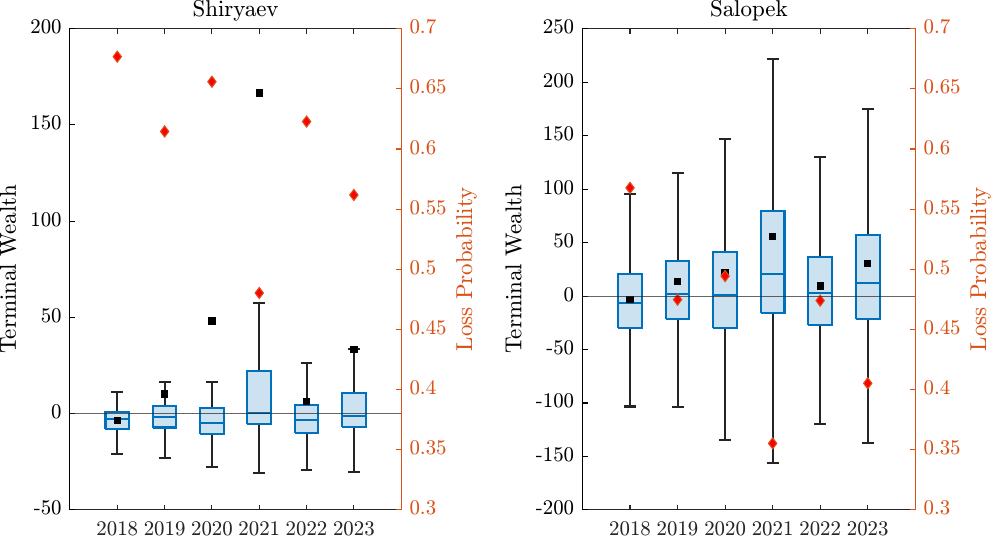}\\ 
				\end{center}	
				\scriptsize
				{\neu For each investment period in our sample, this figure visualizes the performance of the Shiryaev and Salopek strategies for the persistent stocks and stock pairs identifiable via Figure \ref{Fig:EmpHurst}. Recall that we trade daily over one year, use a rebasing to one and set $p = (0.1,0)$, $\gamma = 10^2$ and $(\alpha,\beta) = (-30,30)$. Besides presenting the terminal wealth distributions in the form of Box-Whisker plots, we report the mean terminal value (black rectangles, left axis) and the empirical loss probabilities (red diamonds, right axis) of the strategies.}  
				\caption{{\neu Empirical strategy performance}}	\label{Fig:EmpStrat}
			\end{figure}
		
		Figures \ref{Fig:EmpHurst} and \ref{Fig:EmpStrat} present our results. For each investment period, Figure \ref{Fig:EmpHurst} plots the distribution of Hurst coefficient estimates over all stocks and highlights the ones that we consider appropriate for our strategies. Interestingly, this only applies to a small fraction of stocks. For example, while we can invest in 99 stocks in 2018, we have 127 in 2021. Using these stocks in the Shiryaev and Salopek strategies delivers the outcomes summarized in Figure \ref{Fig:EmpStrat}. It yearly visualizes the terminal wealth distributions of the strategies in the form of Box-Whisker plots. In addition, it provides the mean terminal wealth and the empirical loss probability. For the Shiryaev strategy, we observe that, with the exception of 2018, it generates positive terminal values on average which are accompanied by rather high loss probabilities. In comparison, the Salopek strategy shows the same mean terminal wealth signs but at significantly lower loss probabilities. Both strategies perform best in the year 2021.\footnote{{\neu The outstanding Shiryaev mean in this year is largely caused by about 20 stocks with exceptional strategy performance outside the plot quantile range.}} The apparently time-varying performance of the strategies can be explained by various factors including parameter instability and randomness. Most importantly, our samples may still contain stocks which cannot be adequately described by a fBm.
	}
	
	\section{Conclusion}
	\label{sec:concl}
	In this study, we have revisited the arbitrage strategies of \cite{Shiryaev1998} and \cite{Salopek1998} {\neu because they have a solid theoretical foundation and an elegant intuitive design making them interesting for investment practice.} While the Shiryaev strategy trades only one risky asset and benefits from both rising and falling prices, the most elementary specification of the Salopek strategy trades two risky assets and capitalizes on prices drifting apart. Both strategies have very simple trading rules and rely only on realized prices, which are readily available in today's investment world, so that they can be easily automated in modern algorithmic trading facilities. 
	
	Because these strategies aim at continuous-time trading in a fractional Black-Scholes market, we have transferred them to a discrete-time application and intensively studied their investment performance via Monte Carlo simulation {\neu and formal mathematical approximation}. In conservative settings with independent assets, moderate serial correlation and realistic transaction costs, we show that, even though they can no longer be considered as arbitrage strategies, they exhibit positive terminal values on average and are accompanied by low loss probabilities. This makes them particularly interesting for tail-oriented investors \cite[see][]{Gao2018}. Furthermore, we have revealed several interesting features of the discretized strategies. First, they perform reasonably well even if assets show relatively small persistence. Second, certain limiting cases of the strategies not only max out their performance but further simplify their implied asset positions. Third, time-discretization does not necessarily lead to portfolio values lower than in the continuous-time case. Finally, when adequately scaled, the strategies are useful for short-, medium- and long-term horizons and most advisable at a daily trading frequency. This nicely integrates into the growing literature on the welfare consequences of speeding up transactions in financial markets \cite[see][]{Du2017}.
	
	Our study offers plenty of scope for future research. With respect to theoretical work, it is instructive to introduce an interest-bearing transaction account (with potentially differing rates for borrowing and lending) {\neu and to consider the market impact of large-scale trades \cite[see][]{Guasoni2019,Guasoni2021}.} Furthermore, modeling a negative cross-correlation between risky assets can be considered a fruitful endeavor because it has the potential to increase strategy performance. It also makes the strategies comparable to the domain of pairs trading rules for correlated assets \cite[see][]{Krauss2017,Chen2017}. {\neu Finally, while we have focused on the usefulness of the strategies in their originally proposed form, it is worth investigating whether they can be improved by modifications and how they perform in a direct comparison with related strategies \cite[see][]{Garcin2022,Xiang2024}. Potential modifications include strategy starting points updated during the investment life cycle, early liquidation rules (e.g., closing when a given minimum gain has been achieved), lock-in mechanisms partially securing ongoing profits (e.g., shorting when the price falls below half of the maximum reached) and adapted time grids (e.g., optimal hitting-based rebalancing instead of rigid end-of-period trading).} {\neu As far as empirical work is concerned, we suggest an extension of our brief empirical illustration to a profound analysis of the Shiryaev and Salopek strategies in different asset classes and markets. Besides emerging market equities, currency investments are a particularly promising playground for such an endeavor because exchange rates show pronounced periodic smoothness, memory and predictability \cite[see][]{Karemera2006,Hassan2019}. Similar statements can be made for stable evolutions in (investment grade) corporate bond prices. However, taking this path requires overcoming the obstacle of sound fBm testing. To achieve this, one could sequentially apply test procedures constructed for a given hypothetical $H$ \cite[see][]{Sikora2018}. Alternatively, these techniques might be suitably combined with a pre-estimate or a concurrent estimation of $H$ based on methods more advanced than rescaled range analysis \cite[see][]{Lopez2021}. Only this way it is possible to reliably answer the question of whether an asset qualifies for our memory-based investment strategies.}
	
	
	\bigskip
	\appendix	
	\vspace{0.5cm}
	\noindent\textbf{\large Appendix}\\[-5ex]
	{
		\section{Proofs}
		\label{app:proofs}	
		{\neu
			\subsection{Proof of Proposition \ref{prop_distribution_shiryaev}}\label{prop_distribution_shiryaev_proof}
			\begin{proof} 
				We use expression \eqref{eq:TermValueShiraev} for the portfolio value of the Shiryaev strategy  and  the price model \eqref{eq:BlackScholes}. They yield
				\begin{align}
					V^\Psi_T=\frac{1}{s_0^1}(S_T^1-s_0^1)^2 &= s_0^1 \big(\exp\{\mu T +\sigma B_T^H\}-1\big)^2 =  s_0^1 \big(e^Z-1\big)^2= s_0^1 \big(e^{2Z} -2 e^Z+1\big).
				\end{align}
				Here, $Z=\mu T +\sigma B_T^H$ is a Gaussian random variable with mean $\mu T$ and variance $\sigma^2T^{2H}$. Hence, $e^{2Z}$ and $e^Z$ are lognormally distributed. Taking expectation provides the proposed expression for $\E[V^\Psi_T]$. The proof for $\E[(V^\Psi_T)^2]$ is similar. The variance follows from Steiner's formula. For the cumulative distribution function of $V_T^\Psi$ with $v\ge 0$, we have
				\begin{align}						
					F_{V_T^\Psi} (v) = \P\big(V_T^\Psi\le v\big) &= \P\big(s_0^1 \big(e^Z-1\big)^2\le v\big) =  \P\Big(-\sqrt{v/s_0^1} \le e^Z-1 \le \sqrt{v/s_0^1}\Big) \\
					& = \P\Big(\log\Big\{\Big(1-\sqrt{v/s_0^1}\Big)^+\Big\} \le Z \le \log\Big\{1+\sqrt{v/s_0^1}\Big\}\Big).
				\end{align}
				Since $Z\sim\mathcal{N}(u,s^2)$ with $u=\mu T$ and $s=\sigma T^H$, the claim follows from $\P(a\le Z\le b)= \overline \Phi((b-u)/s) - \overline \Phi((a-u)/s)$ for $a\le b$. \qed
			\end{proof}
		}
		
		{\neu
			\subsection{Proof of Lemma \ref{lemma}}\label{lemma_proof}
			\begin{proof}
				Based on the transaction account recursion \eqref{eq:rebal_strat}, we infer \eqref{def_cum_rebalancing} from \eqref{eq:value_cont_discrete}. To show that \eqref{def_cum_rebalancing} also holds at liquidation, we use \eqref{eq:rebal_strat} and the revenue expression \eqref{eq:revenue_liq} to obtain
				
				\begin{align}
					\label{termvalue1}
					V_{T}^{\stratd}&= R^{\stratd} +{\stratd}_{N}^{d+1} =  \sum_{i=0}^{d}\stratd_N^i S_{T}^i - \sum_{n=1}^{N-1} D^{\stratd}_{t_n} - \transactcosts.
				\end{align}		
				Taking into account that $\stratd_N=\stratc_{t_{N-1}}$, we have
				\begin{align}
					\label{dec}
					\sum_{i=0}^{d}\stratd_N^i S_{T}^i  = \sum_{i=0}^{d}\stratc_{t_{N-1}}^i S_{T}^i =  \sum_{i=0}^{d}\stratc_{T}^iS_{T}^i - D^{\stratd}_{t_N} = V_T^\stratc - D^{\stratd}_{t_N}, 
				\end{align}	
				where $D^{\stratd}_{t_N} = \sum_{i=0}^{d} \big(\Psi_{t_N}^i-\Psi_{t_{N-1}}^i\big)S_{T}^i$ follows the rebalancing cost definition $D^{\stratd}_{t_n}$, $n=1,\ldots,N-1$, in  \eqref{eq:Difference} but is now evaluated at terminal time $t_N=T$. Because there is no further rebalancing at time $T$, these costs may be called ``virtual costs''. They are not booked to the transaction account. Substituting \eqref{dec} into \eqref{termvalue1} yields	
				$V_T^\stratc -V_{T}^{\stratd}= \rebalcosts  + \transactcosts$ and proves the validity of \eqref{def_cum_rebalancing} for $t=T$.
				Here, $ \rebalcosts$ is defined as in \eqref{def_cum_rebalancing}. It represents the  cumulated rebalancing costs over the entire trading period $[0,T]$ including the virtual cost term $D^{\stratd}_{t_N}$. \qed
			\end{proof}
		}
		
		\subsection{Proof of Proposition \ref{prop_Salopek}}\label{prop_Salopek_proof}
		\begin{proof}
			Recall the Salopek strategy \eqref{eq:continuousSalopekStrategy} where 
			$\stratc_{t}^i(\alpha,\beta)= \widehat\stratc_t^i(\beta)-\widehat\stratc_t^i(\alpha)$, 
			$\widehat{\stratc}_t^i(a)= \frac{1}{d}\left(\frac{S_t^i}{M_a(S_t)}\right)^{a-1}$, and $M_a(x)$ is the \eqref{eq:alphaorder} $a$-order power mean of $x=(x^1,\ldots,x^d)\in\R_+^d$. 
			
				We start with proving $\stratc_t^i(\alpha,\beta) >0$.
			Denoting $x={S_t^i}/{M_\beta(S_t)}$ and $y={S_t^i}/{M_\alpha(S_t)}$, relation \eqref{eq:Salopek_monoton} implies $1 <x<y$ and $\beta\ge 1> \alpha$ gives $x^{\beta-1}\ge 1 >y^{\alpha-1}$. Hence, $({S_t^i}/{M_\beta(S_t)})^{\beta-1} >({S_t^i}/{M_\alpha(S_t)})^{\alpha-1}$. This proves  $\widehat\stratc_t^i(\beta) > \widehat\stratc_t^i(\alpha)$ from which the claim follows.
			
			The proof of $\stratc_t^j(\alpha,\beta) <0$ is similar. 
			Denoting $x={S_t^j}/{M_\beta(S_t)}$ and $y={S_t^j}/{M_\alpha(S_t)}$, relation \eqref{eq:Salopek_monoton} implies $0<x<y<1$ and $\beta\ge 1> \alpha$ gives $x^{\beta-1}\le 1 < y^{\alpha-1}$. Hence, $({S_t^j}/{M_\beta(S_t)})^{\beta-1} <({S_t^j}/{M_\alpha(S_t)})^{\alpha-1}$. This proves  $\widehat\stratc_t^j(\beta) < \widehat\stratc_t^j(\alpha)$ from which the claim follows.
			
			For the special case $S^i_t=S_t^{max}=\max\{S_t^1,\ldots S_t^d\}$ and $S^j_t=S_t^{min}=\min\{S_t^1,\ldots S_t^d\}$, the monotonicity property \eqref{eq:mean_monoton} of the $a$-order power mean $M_a(S_t)$ saying that $S_t^{min}\le M_\alpha(S_t)\le M_\beta(S_t)\le S_t^{ max} $    implies \eqref{eq:Salopek_monoton}. This completes the proof. \qed		
		\end{proof}
		
		{\neu		
			\subsection{Proof of Proposition \ref{Shiryaev_rebalancing}}\label{Shiryaev_rebalancing_proof}
			\begin{proof}
				Substituting the Shiryaev strategy \eqref{eq:Shiryaev_stratc} into \eqref{eq:Difference} and using $\Phi_n^i=\Psi_{t_{n-1}}^i, n=1,\ldots, N,$ we have almost surely
				\begin{align}
					D_{t_n}^{\stratd}&=\frac{(S_{t_{n-1}}^1)^2-(S_{t_n}^1)^2}{s_0^1}S^0_{t_n}+\frac{2}{s_0^1}\big(S_{t_n}^1-S_{t_{n-1}}^1\big)S^1_{t_n} \\
					&=\frac{1}{s_0^1}\big((S_{t_n}^1)^2+(S_{t_{n-1}}^1)^2-2S_{t_n}^1S_{t_{n-1}}^1\big)
					=\frac{(S_{t_n}^1-S_{t_{n-1}}^1)^2}{s_0^1}>0. \tag*{\qed}		
				\end{align}	
			\end{proof}
		}
		
		{\neu
			\subsection{Proof of Proposition \ref{prop_conv_shiryaev}}\label{prop_conv_shiryaev_proof}	
			\begin{proof} 
				Based on Proposition \ref{Shiryaev_rebalancing}, i.e., expression \eqref{rebalance_shiryav} for the rebalancing costs $ D_{t_n}^{\stratd}$ of the Shiryaev strategy, the cumulative rebalancing costs are given by
				\begin{align} 
					\label{cum_rebal_costs_shiryaev}
					\rebalcosts(\Delta t)& =\sum_{n=1}^{N} D_{t_n}^{\stratd}
					=\frac{1}{s_0^1} \sum_{n=1}^{N}  \hshir(t_{n-1},\Delta t)~~ 
					\text{with}~~\hshir(t,\Delta t)  =(S_{t+\Delta t}^1-S_{t}^1)^2,~t\in[0,T-\Delta t].		
				\end{align}		
				We now analyze the asymptotic behavior of $ \hshir(t,\Delta t)$ for $\Delta t \to 0$. Applying the Taylor expansion $e^x=1+x +O(x^2)$ for $x\to 0$ to the increments of the price process  \eqref{eq:BlackScholes}, we obtain
				\begin{align*}
					S_{t+\Delta t}^1-S_{t}^1 &= S_{t}^1\big(\exp\big\{\mu\Delta t +\sigma\Delta B^H_t\big\}  -1\big) \\
					&=  S_{t}^1\big(\mu\Delta t +\sigma\Delta B^H_t +O(\Delta t^2) + O((\Delta B^H_t)^2)+  O(\Delta t\Delta B^H_t)\big),
				\end{align*}
				where we have introduced the shorthand notation $\Delta B^H_t = B_{t+\Delta t}^H-B_t^H$.
				Substituting this expansion into \eqref{cum_rebal_costs_shiryaev}, taking expectation, and considering that $\Delta B^H_t\sim\mathcal{N}(0,\Delta t^{2H} )$ implies $\E\big[(\Delta B^H_t)^2\big] =\Delta t^{2H}$ and  $\E\big[(\Delta B^H_t)^k\big] = o(\Delta t^{2H})$ for $k\ge 3$, it follows that	
				\begin{align} 
					\label{hh1}
					\E\big[\hshir(t,\Delta t) \big]
					&=   \E \big[ e^{2\sigma B_{t}^H}(\Delta B^H_t)^2\big]\, 
					\sigma^2(s^1_0)^2e^{2\mu t}+o(\Delta t^{2H}).
				\end{align}
				Here, we have used the boundedness of the moments of $S^1_{t}=s_0^1\exp\{\mu t+\sigma B_{t}^H\}$.
				The expectation on the r.h.s.~of \eqref{hh1} can be rewritten as $\E [ e^{Z_1}Z_2^2]$, where   $Z_1=2\sigma B_{t}^H$ and $Z_2=\Delta B^H_t$  form a zero-mean Gaussian random vector $(Z_1,Z_2)^\top$ with variances $\E[Z_1^2]=4\sigma^2t^{2H}$ and $\E[Z_2^2]=\Delta t^{2H}$. For the co\-variance, \eqref{fBm_covariance} implies
				\begin{align}					
					\begin{split}
						\label{hh2}
						\Cov(Z_1,Z_2) 
						= 2\sigma \Cov(B_{t}^H, B_{t+\Delta t}^H-B_{t}^H) & = \sigma \big((t+\Delta t)^{2H} -t^{2H} -\Delta t^{2H}\big)\\
						& = 2\sigma Ht^{2H-1} \Delta t+ O(\Delta t^{2H}).	
					\end{split}					
				\end{align}
				Applying Isserli's theorem for the computation of the moments of a zero-mean Gaussian vector \cite[see][]{Michalowicz2009}, some algebra delivers, for the r.h.s. expectation of \eqref{hh1}, the expression
				\begin{align}
					\E \big[ e^{2\sigma B_{t}^H}(\Delta B^H_t)^2\big] = \E \big[ e^{Z_1}Z_2^2\big] &= \big(\E\big[Z_2^2] +(\Cov(Z_1,Z_2))^2 \big)\E \big[ e^{Z_1}]\\
					&=  \big(\Delta t^{2H} + O(\Delta t^2)\big) \E \big[ e^{2\sigma B_{t}^H}] =  \exp\big\{2\sigma^2 t^{2H}\big\} \Delta t^{2H} + O(\Delta t^2),
				\end{align}
				where we have taken into account the expansion \eqref{hh2} for $\Cov(Z_1,Z_2)$, and that $e^{2\sigma B_{t}^H}$ is lognormally distributed with mean $\exp\{2\sigma^2 t^{2H}\}$. Substituting into \eqref{hh1} and \eqref{cum_rebal_costs_shiryaev} yields
				\begin{align*} 
					\E\big[\rebalcosts(\Delta t) \big]
					&= \frac{1}{s_0^1} \sum_{n=1}^{N}  \Big(\sigma^2(s^1_0)^2e^{2\mu t_{n-1}}\exp\big\{2\sigma^2 t_{n-1}^{2H}\big\} \Delta t^{2H} \; +o(\Delta t^{2H})\Big)\\
					& = {s_0^1} \sigma^2 \,\frac{1}{\Delta t}\Big(\int_0^T  \exp\big\{2 \mu t + 2\sigma^2 t^{2H}\big\} \,dt +O(\Delta t)\Big)\, \Big(\Delta t^{2H} +o(\Delta t^{2H})\Big).		
				\end{align*}
				In this context, we have exploited the linear approximation order of the composite rectangular rule applied to the  integral $\int_0^T  \exp\{2 \mu t + 2\sigma^2 t^{2H}\} \,dt$.
				Multiplying out the brackets proves the claim. \qed		 
				
			\end{proof}
		}
		
		{\neu
			\subsection{Proof of Proposition \ref{prop_conv_salopek}}\label{prop_conv_salopek_proof}	
			\begin{proof} 
				Recall the Salopek strategy definition in \eqref{eq:continuousSalopekStrategy}. We have $\Psi_{t}^i=\Psi_{t}^i(\alpha,\beta)= \widehat\Psi_t^i(\beta)-\widehat\Psi_t^i(\alpha)$, $i=1,\ldots,d$, where 
				$\widehat{\Psi}_t^i(a)		=\frac{1}{d}\big(\frac{S_t^i}{M_a(S_t)}\big)^{a-1}$, and $M_a(S_t)$ is the $a$-order power mean of $S_t$. Using expression \eqref{eq:Difference}  for the rebalancing costs $ D_{t_n}^{\stratd}$, the cumulative rebalancing costs are given by
				\begin{align} 
					\begin{split}
						\label{cum_rebal_costs_salopek}
						\rebalcosts(\Delta t)& =\sum_{n=1}^{N} D_{t_n}^{\stratd} 
						= \sum_{n=1}^{N}  \sum\limits_{i=1}^{d} \big(\Psi_{t_n}^i-\Psi_{t_{n-1}}^i\big)S_{t_n}^i	=	\sum_{n=1}^{N} \hsalop(t_{n-1},\beta,\Delta t)-  \hsalop(t_{n-1},\alpha,\Delta t) 	\\
						\text{with} ~ \hsalop(t,a,\Delta t) & = \sum\limits_{i=1}^{d} \big(\widehat{\Psi}^i_{t+\Delta t}(a) - \widehat{\Psi}^i_t(a)\big) S^i_{t+\Delta t}, \quad t\in[0,T-\Delta t],~ a\in\R.
					\end{split}
				\end{align}		
				We now analyze the asymptotic behavior of $ \hsalop(t,a,\Delta t)$ for $\Delta t \to 0$. Applying the Taylor expansion $e^x=1+x + x^2/2 +O(x^3)$ for $x\to 0$ to the price process \eqref{eq:BlackScholes} yields
				\begin{align}
					\label{S_expansion}
					\frac{S^i_{t+\Delta t}}{S^i_t} &= \exp\{\mu^i \Delta t +\sigma^i(B_{t+\Delta t}^{H^i}-B_t^{H^i})\} \\		
					&= 1+ \mu^i \Delta t +\sigma^i \Delta B^{H^i}_t + \frac{1}{2} (\sigma^i)^2(\Delta B^{H^i}_t)^2 +O(\Delta t^2) +O(\Delta t \Delta B^{H^i}_t),
				\end{align}
				where we have introduced the shorthand notation $\Delta B^{H^i}_t = B_{t+\Delta t}^{H^i}-B_t^{H^i}$. It holds that
				\begin{align} 		
					\hsalop(t,a,\Delta t) & = \sum\limits_{i=1}^{d} \widehat{\Psi}^i_{t+\Delta t}(a)S^i_{t+\Delta t} - \sum\limits_{i=1}^{d} \widehat{\Psi}^i_t(a) S^i_{t} \frac{S^i_{t+\Delta t}}{S^i_t} \\
					&=~~  \frac{1}{d M^{a-1}_a(S_{t+\Delta t})} \sum\limits_{i=1}^{d} (S_{t+\Delta t}^i)^{a-1} \, S^i_{t+\Delta t}\\
					&~~~~~			-~\frac{1}{d M^{a-1}_a(S_{t})} ~~ \sum\limits_{i=1}^{d} (S_{t}^i)^{a-1} \, S^i_{t}\Big(1+   \frac{1}{2} (\sigma^i)^2(\Delta B^{H^i}_t)^2\Big) - \mathcal{R}(t,a,\Delta t)\\
					&= M_a(S_{t+\Delta t}) - M_a(S_{t})  - \frac{1}{2d M^{a-1}_a(S_{t})} \sum\limits_{i=1}^{d} (S_{t}^i)^{a}\, (\sigma^i)^2(\Delta B^{H^i}_t)^2 - \mathcal{R}(t,a,\Delta t),
					\label{hh3}
				\end{align}		
				where the remainder term is given by 
				\begin{align} 
					\label{Rdef}		
					\mathcal{R}(t,a,\Delta t) & =
					\frac{1}{d M^{a-1}_a(S_{t})} \sum\limits_{i=1}^{d} (S_{t}^i)^{a}\Big(\mu^i \Delta t +\sigma^i \Delta B^{H^i}_t+O(\Delta t^2) +O(\Delta t \Delta B^{H^i}_t)\Big).
				\end{align}		
				Next, we derive an asymptotic expansion of the first term in \eqref{hh3}, i.e.,  the power mean $M_a(S_{t+\Delta t})$, around the point $S_{t}$ using the Taylor expansion in \eqref{S_expansion}.  We have
				\begin{align} 		
					M_a(S_{t+\Delta t})  & = \bigg( \frac{1}{d}\sum\limits_{i=1}^{d} (S_{t+\Delta t}^i)^{a}\bigg)^{1/a}	=
					\bigg( \frac{1}{d}\sum\limits_{i=1}^{d} (S_{t}^i)^{a} + \Delta Y\bigg)^{1/a} = \big( M_a^a(S_t) + \Delta Y\big)^{1/a}	 \\
					\text{with}~~\Delta Y& = \frac{1}{d}\sum\limits_{i=1}^{d} (S_{t}^i)^{a} \Big(a\mu^i \Delta t +a\sigma^i \Delta B^{H^i}_t + \frac{a^2}{2} (\sigma^i)^2\big(\Delta B^{H^i}_t\big)^2 +O(\Delta t^2) +O\big(\Delta t \Delta B^{H^i}_t\big)\Big).	
				\end{align}			
				The second-order Taylor expansion of the function $x\mapsto f(x)=x^{1/a}$ defined on $(0,\infty)$ around the point $x=x_0>0$ 
				\begin{align}
					f(x_0+\Delta x) &= x_0^{1/a}\Big(1+\frac{1}{ax_0} \Delta x + \frac{1-a}{2a^2 x_0^2} \Delta x^2 + O(\Delta x ^3)\Big)
				\end{align}
				provides, with $x_0=M_a^a(S_t)$ and $\Delta x=\Delta Y$, the expression
				{\footnotesize
					\begin{align}
						M_a(S_{t+\Delta t})  = M_a(S_{t}) &+	\frac{1}{d M^{a-1}_a(S_{t})} \sum\limits_{i=1}^{d} (S_{t}^i)^{a}\Big(\mu^i \Delta t +\sigma^i \Delta B^{H^i}_t + \frac{a}{2} (\sigma^i)^2(\Delta B^{H^i}_t)^2 +O(\Delta t^2) +O(\Delta t \Delta B^{H^i}_t)\Big)\\
						& + \frac{1-a}{2d^2 M^{2a-1}_a(S_{t})} \sum\limits_{i,j=1}^{d} (S_{t}^iS_{t}^j)^{a}\Big(\sigma^i \sigma^j \Delta B^{H^i}_t\Delta B^{H^j}_t +\mathcal{J}\Big),
					\end{align}
				}
				where $\mathcal{J}$ collects all the remainder terms which, after taking expectation (see below), are of order $o(\Delta t^{2H})$.
				Substituting into \eqref{hh3}, considering  the definition of $\mathcal{R}(t,a,\Delta t)$ in \eqref{Rdef}, and taking expectation supplies
				\begin{align} 		
					\E[\hsalop(t,a,\Delta t)] & = \frac{a-1}{2d} \sum\limits_{i=1}^{d} (\sigma^i)^2 \E\bigg[\frac{(S_{t}^i)^{a}}{M^{a-1}_a(S_{t})}\bigg(1-\frac{1}{d}\Big(\frac{S_t^i}{M_a(S_t)}\Big)^a\bigg)(\Delta B^{H^i}_t)^2\bigg]  +o(\Delta t^{2H}).
				\end{align}		
				Here, we have used the independence of the fBms $B^{H^1},\ldots,B^{H^d}$ which implies that $\E[\Delta B^{H^i}_t\Delta B^{H^j}_t]=0$ for $i\neq j$. An application of Isserli's theorem similar to the proof of Proposition \ref{prop_conv_shiryaev} in Appendix \ref{prop_conv_shiryaev_proof} shows that all ``mixed terms'' ($i\neq j$) in the double sum do not contribute to the leading term of the expansion.
				Another application of the theorem to the expectation on the r.h.s. yields
				\begin{align} 		
					\E[\hsalop(t,a,\Delta t)] & = \frac{a-1}{2d} \sum\limits_{i\in \Hminindex} (\sigma^i)^2 \E\bigg[\frac{(S_{t}^i)^{a}}{M^{a-1}_a(S_{t})}\bigg(1-\frac{1}{d}\Big(\frac{S_t^i}{M_a(S_t)}\Big)^a\bigg)\bigg]\Delta t^{2H_{\min}}  +o(\Delta t^{2H_{\min}}),
				\end{align}		
				where we utilize the fact that the leading term of the expansion only collects the terms which are  related to assets with  the smallest Hurst coefficients, i.e.,  $H^i=H_{\min}$.
				In view of \eqref{cum_rebal_costs_salopek}, we have to evaluate sums of the form  $F=\sum_{n=1}^{N} \E[\hsalop(t_{n-1},a,\Delta t)]	$ for which we obtain 
				\begin{align} 				
					F	
					& = \sum_{n=1}^{N} \frac{a-1}{2d}  \sum\limits_{i\in \Hminindex} \Bigg\{(\sigma^i)^2
					\E\bigg[\frac{(S_{t_{n-1}}^i)^{a}}{M^{a-1}_a(S_{t_{n-1}})} \bigg(1-\frac{1}{d}\Big(\frac{S_{t_{n-1}}^i}{M_a(S_{t_{n-1}})}\Big)^a\bigg)\bigg]  \Delta t^{2H_{\min}}
					+o(\Delta t^{2H_{\min}})\Bigg\}\\
					&= \frac{a-1}{2d}  \frac{1}{\Delta t} \Bigg\{   \int_0^T 
					\E\bigg[M_a(S_{t})\sum\limits_{i\in \Hminindex} (\sigma^i)^2 \Big(\frac{S_{t}^i}{M_a(S_{t})}\Big)^a \bigg(1-\frac{1}{d}\Big(\frac{S_{t}^i}{M_a(S_{t})}\Big)^a\bigg)\bigg]\,dt +O(\Delta t)\Bigg\} \times\\
					& \hspace*{20em} \times \big(\Delta t^{2H_{\min}}+o(\Delta t^{2H_{\min}})\big)\\
					&= \overline{C}(a) \Delta t^{2H_{\min}-1} + o(\Delta t^{2H_{\min}-1}).
				\end{align}		
				In this formulation, we consider the linear approximation order of the composite rectangular rule applied to the  integral in the second line, and the definition of $\overline{C}(a)$ in the proposition.
				Substituting the above expression into   \eqref{cum_rebal_costs_salopek} leads to	{\footnotesize					
					\begin{equation} 				
						\E[\rebalcosts(\Delta t)]=	\sum_{n=1}^{N} \big(\E[\hsalop(t_{n-1},\beta,\Delta t)]-  \E[\hsalop(t_{n-1},\alpha,\Delta t)]\big)		
						= (\overline{C}(\beta) -\overline{C}(\alpha))\Delta t^{2H_{\min}-1}  +o(\Delta t^{2H_{\min}-1}),
					\end{equation}
				}
				which proves the claim. 	\ 	\qed
			\end{proof} 
		}
		
		\section{Additional results}

		Figures \ref{Fig:SalopekPositiveDifferences} and \ref{Fig:SalopekStrategyLimit} supply additional simulated realizations of the Salopek strategy to illustrate the effects of negative rebalancing costs and infinite strategy parameter values, respectively.
		
			\begin{figure}[tb!h]
				\begin{center}
					\includegraphics[trim= 10mm 10mm 0mm 10mm, clip, width=11.25cm]{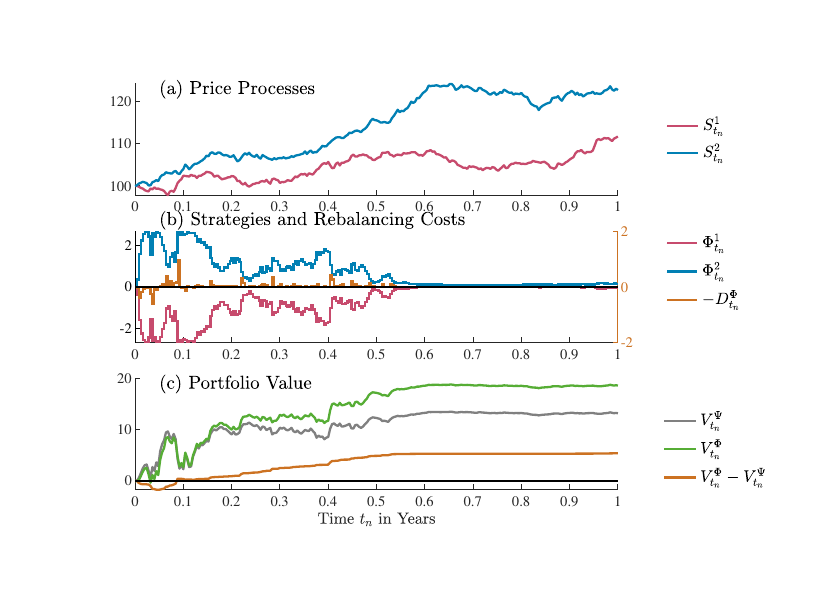}
				\end{center}	
				\scriptsize
				Using the overall design and setting of Figure \ref{Fig:Salopek1}, this figure simulates another exemplary realization of the Salopek strategy and replaces the $(\alpha,\beta)$ basis values of $(-30,30)$ by $(71,80)$.
				\caption{Realization of the discretized Salopek strategy with negative rebalancing costs}	
				\label{Fig:SalopekPositiveDifferences}
			\end{figure}
			
			\begin{figure}[tbh]
				\begin{center}
					\includegraphics[trim= 10mm 10mm 0mm 10mm, clip, width=11.25cm]{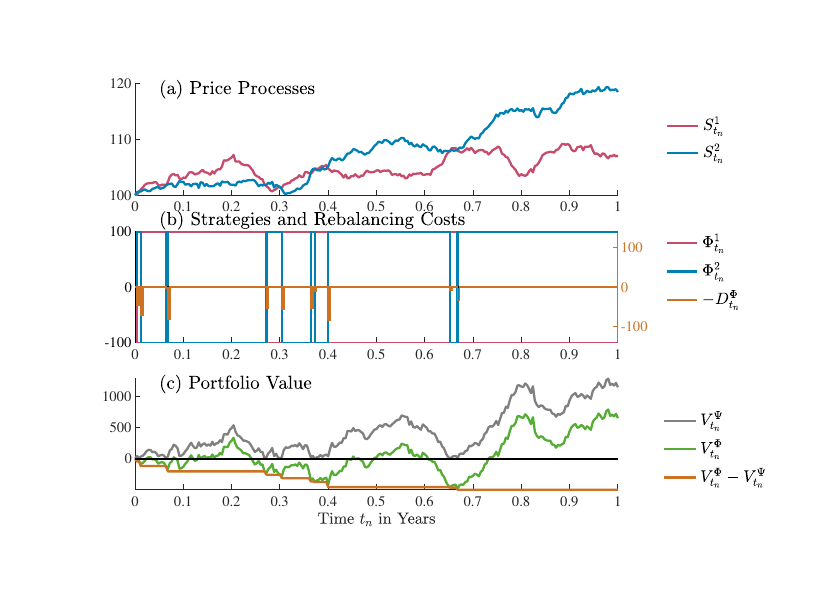}
				\end{center}	
				\scriptsize
				Using the simulated prices of Figure \ref{Fig:Salopek1}, this figure illustrates the behavior of the Salopek strategy in a situation where the $(\alpha,\beta)$ basis values of $(-30,30)$ are replaced by $(-\infty,\infty)$.
				\caption{Realization of the discretized Salopek strategy with parameter infinity}
				\label{Fig:SalopekStrategyLimit}
			\end{figure}
		
		{\neu 
			\section{Alternative price model}\label{sec:mfBm}
			
			In our main analysis, we have assumed that asset prices follow the fBm defined in \eqref{eq:BlackScholes}. A straightforward alternative price model is the mixed fBm (mfBm). With simplified notation, i.e., after dropping the asset index $i$, it is given by
			\begin{equation}\label{eq:mfBm}
				S_t=s_0\, \exp\big\{ \mu t+\nu W_t-\frac{1}{2}\nu^2 t + \sigma B_t^{H}\big\},
			\end{equation}
			where $W_t$ is a sBm and $B_t^{H}$ represents a fBm with $H>0.5$. Furthermore, $W_t$ and $B_t^{H}$ are independent and $s_0,\sigma,\nu >0$, $\mu \in \mathbb{R}$ \cite[see][]{Bender2011}. 
			
			The fBm and the mfBm basically have the same covariance structure. That is, for $H>0.5$, they both exhibit long range dependence controlled by $H$. However, the models differ with respect to their quadratic variation. While, in the case of $H>0.5$, a fBm has zero quadratic variation, a mfBm displays the same positive quadratic variation as the sBm. This is important because our continuous-time arbitrage strategies require that the stock price model features zero quadratic variation \cite[see][]{Bender2008,Bender2011}.
			
			To see how our discretized strategies perform in the presence of positive quadratic variation, we repeat the simulations related to our basis setting (see Table \ref{Tab:paramters}) by generating mfBm prices instead of fBm prices. We start with a discussion of the merits of the Shiryaev strategy in a mfBm environment with $\nu=0.05$. Following the style of Figure \ref{Fig:Shiryaev2}, Figure \ref{Fig:mixShiryaev} compares $\nu=0$ (fBm) to $\nu=0.05$ (mfBm). It shows that, even though the introduction of a small sBm component destroys the strategy's continuous-time arbitrage property, the discretized version of the trading rule can be useful because it generates a positive terminal value on average and extends the range of potential gains. However, as substantiated by the descriptive statistics of Table \ref{Tab:Shiryaev_nu}, this comes at the cost of a higher loss probability. The additional $\nu$ values covered by Table \ref{Tab:Shiryaev_nu} reveal that both the terminal mean and the loss probability rise with $\nu$. The latter increases drastically.
			
			\afterpage{
				\begin{figure}[ht!]
					\begin{center}
						\includegraphics[trim= 0mm 46mm 0mm 10mm, clip, width=11cm]{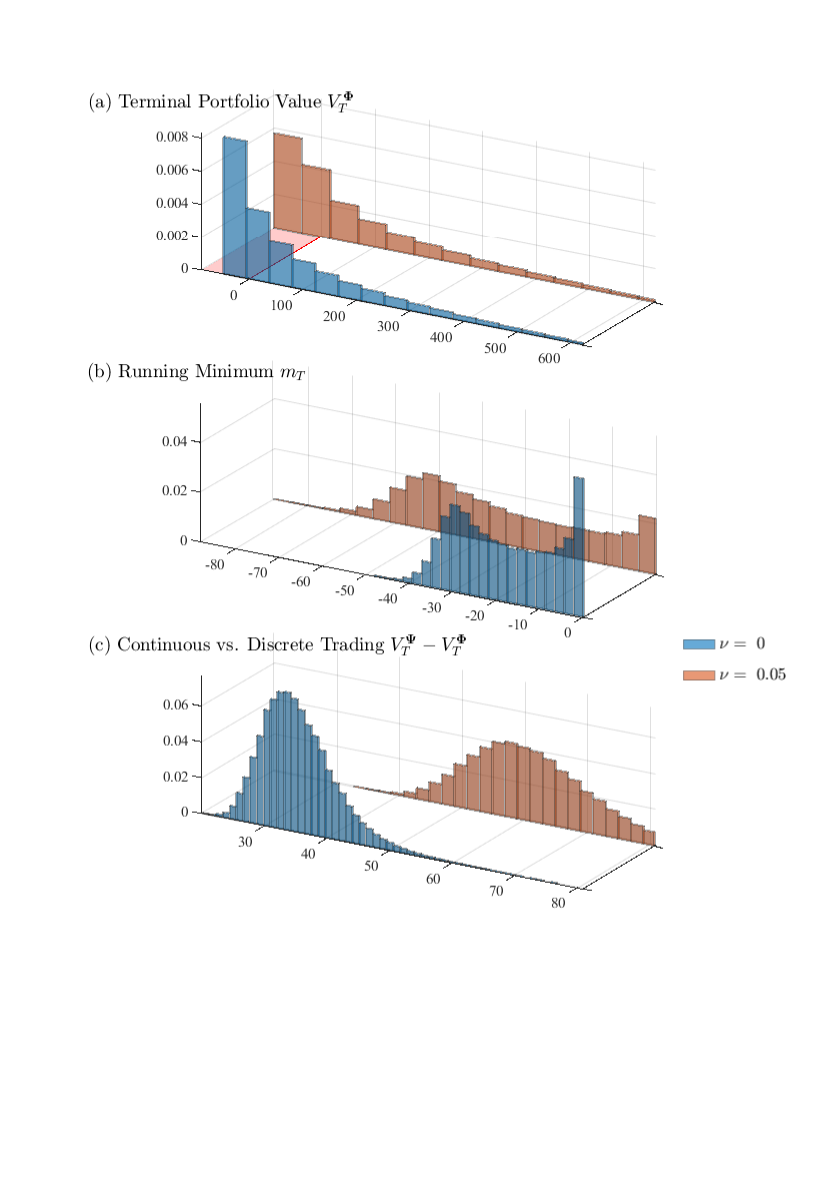}
					\end{center}		
					\scriptsize  
					{\neu This figure reevaluates the basis setting results of the discretized Shiryaev strategy in Figure \ref{Fig:Shiryaev2} by introducing a new price model, i.e., the mfBm specified in \eqref{eq:mfBm}. While $\nu=0.05$ resembles a genuine mfBm, $\nu=0$ reduces the mfBm to a fBm. For better visibility, the x-axis in Panels (a) and (b) ends at the 95\% quantile of $\nu=0.05$.}
					\caption{{\neu Shiryaev portfolio value distributions for alternative price model}}
					\label{Fig:mixShiryaev}
				\end{figure}
				
				\begin{table}[h!]
					\footnotesize
					\begin{center}
						{\neu 		
							\begin{tabular}{rr|rrrrrr}
								\toprule
								$\nu$& $H$ &  Mean & Median & Stand. dev.  & Min & Max & Loss prob. \\\midrule
								0.00 & 0.6 & 107.9 & 25.6 & 218.1 & $-$48.2 & 4,148.5 & 0.39 \\
								0.05 & 0.6 & 111.3 & 11.6 & 266.5 & $-$87.8 & 4,567.0 & 0.46 \\
								0.10 & 0.6 & 114.8 & $-$31.7 & 406.2 & $-$213.1 & 7,234.5 & 0.56 \\
								0.15 & 0.6 & 120.9 & $-$104.2 & 651.1 & $-$449.4 & 15,600.4 & 0.62 \\
								0.20 & 0.6 & 129.5 & $-$203.4 & 1,020.8 & $-$839.0 & 31,797.1 & 0.65 \\
								
								\midrule
								0.00 & 0.9 & 141.3 & 58.3 & 217.9 & $-$1.8 & 3,764.9 & 0.06 \\
								0.05 & 0.9 & 145.3 & 44.7 & 268.7 & $-$38.4 & 4,985.2 & 0.32 \\
								0.10 & 0.9 & 149.3 & 0.5 & 409.7 & $-$146.9 & 7,299.8 & 0.50 \\
								0.15 & 0.9 & 156.1 & $-$72.5 & 657.0 & $-$362.9 & 13,721.6 & 0.59 \\
								0.20 & 0.9 & 165.7 & $-$171.9 & 1,030.3 & $-$691.4 & 26,376.8 & 0.63 \\
								
								\bottomrule
							\end{tabular}
						}
					\end{center}
					\scriptsize
					{\neu Supplementing Figure \ref{Fig:mixShiryaev}, this table reports some descriptive statistics for the terminal portfolio value distributions of the discretized Shiryaev strategy under the new price model \eqref{eq:mfBm} with $\nu \in \{0, 0.05, 0.1, 0.15, 0.2 \}$ and $H \in \{0.6, 0.9\}$.}
					\caption{{\neu \label{Tab:Shiryaev_nu} Shiryaev portfolio value statistics for alternative price model}}
				\end{table}
			}
			
			Turning to the Salopek strategy, Figure \ref{fig:mixSalopek} and Table \ref{Tab:Salopek_nu} indicate that, similar to the Shiryaev case, a mfBm increases the range of positive and negative terminal portfolio values. However, for the Salopek rule, a growing $v$ lowers the terminal mean while simultaneously raising the loss probability. Nevertheless, in the considered $v$ range, it remains beneficial for investors.
			
			\afterpage{
				\begin{figure}[ht!]
					\begin{center}
						\includegraphics[trim= 0mm 46mm 0mm 10mm, clip, width=11cm]{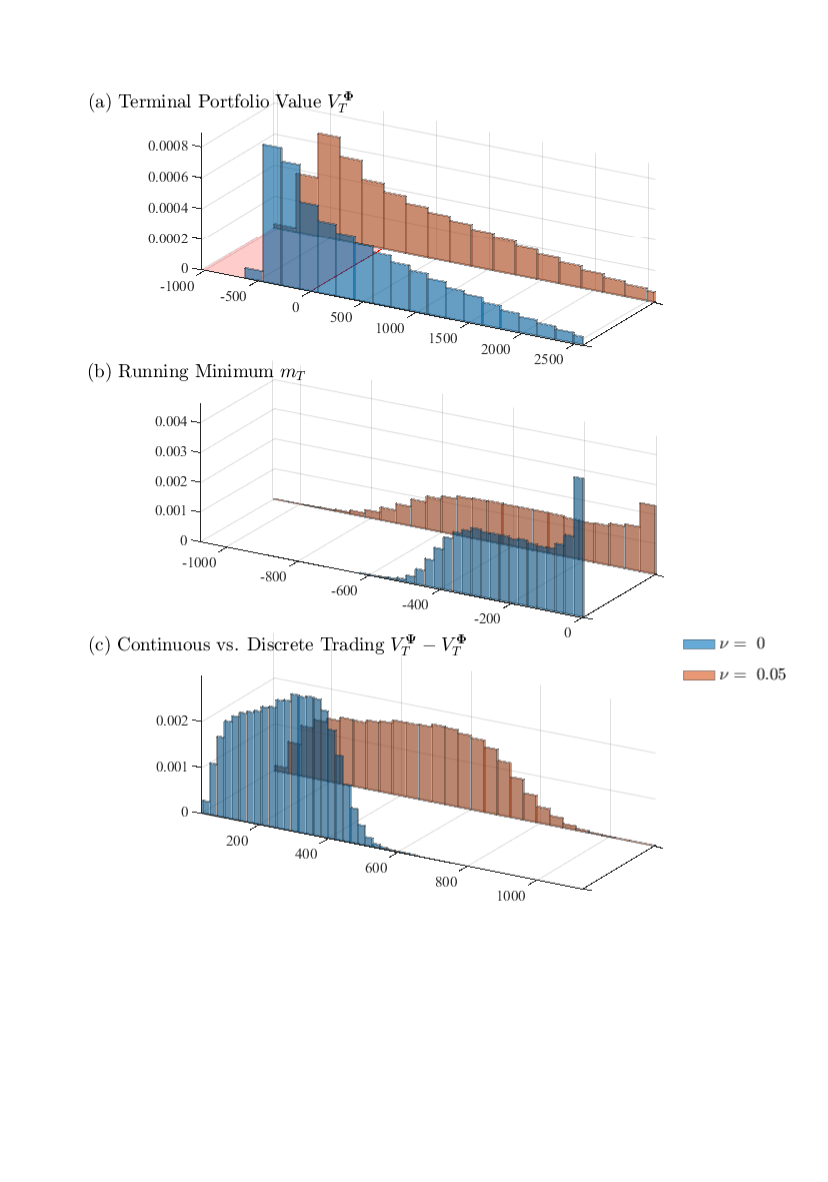}
					\end{center}
					\scriptsize
					{\neu This figure reevaluates the basis setting results of the discretized Salopek strategy in Figure \ref{Fig:Salopek2} by introducing a new price model, i.e., the mfBm specified in \eqref{eq:mfBm}. While $\nu=0.05$ resembles a genuine mfBm, $\nu=0$ reduces the mfBm to a fBm. For better visibility, the x-axis in Panel (a) ends at the 95\% quantile of $\nu=0.05$.}
					\caption{{\neu Salopek sensitivity to strategy parameters for alternative price model}}
					\label{fig:mixSalopek}
				\end{figure}
				
				\begin{table}[h!]
					\footnotesize
					\begin{center}
						{\neu 
							\begin{tabular}[t]{rr|rrrrrr}
								\toprule
								$\nu$& $H$ &  Mean & Median & Stand. dev.  & Min & Max & Loss prob. \\\midrule
								
								0.00 & 0.6 & 533.4 & 277.8 & 887.5 & $-$625.6 & 6,454.2 & 0.37 \\
								0.05 & 0.6 & 495.4 & 215.7 & 1,056.7 & $-$1,027.4 & 7,478.2 & 0.42 \\
								0.10 & 0.6 & 424.2 & 104.9 & 1,490.1 & $-$2,313.9 & 9,766.3 & 0.47 \\
								0.15 & 0.6 & 365.3 & 15.4 & 2,071.9 & $-$4,091.7 & 12,395.4 & 0.50 \\
								0.20 & 0.6 & 326.6 & $-$58.3 & 2,731.2 & $-$7,063.0 & 16,219.1 & 0.51 \\
								\midrule
								0.00 & 0.9 & 794.7 & 547.8 & 810.7 & $-$18.2 & 6,752.4 & 0.07 \\
								0.05 & 0.9 & 717.3 & 443.4 & 995.5 & $-$509.3 & 7,705.6 & 0.31 \\
								0.10 & 0.9 & 584.0 & 268.9 & 1,449.4 & $-$1,855.2 & 9,813.8 & 0.43 \\
								0.15 & 0.9 & 480.9 & 130.2 & 2,044.6 & $-$3,780.2 & 12,080.3 & 0.48 \\
								0.20 & 0.9 & 413.7 & 26.7 & 2,712.4 & $-$6,384.2 & 16,651.3 & 0.50 \\
								
								\bottomrule
							\end{tabular}
						}
					\end{center}
					\scriptsize
					{\neu Supplementing Figure \ref{fig:mixSalopek}, this table reports some descriptive statistics for the terminal portfolio value distributions of the discretized Salopek strategy under the new price model \eqref{eq:mfBm} with $\nu \in \{0, 0.05, 0.1, 0.15, 0.2 \}$ and $H \in \{0.6, 0.9\}$.}
					\caption{{\neu Salopek portfolio value statistics for alternative price model}}
					\label{Tab:Salopek_nu} 
				\end{table}
			}
		} 
		
		
		\newpage		
		\clearpage
		
		
		
		{\footnotesize

		}

	\end{document}